\documentclass[a4paper,12pt]{article}

\usepackage{amsmath,amssymb,mathtools} 
\usepackage{color}
\usepackage{graphicx}
\usepackage{subfigure}
\usepackage{cite}           
\usepackage{hyperref}            
\usepackage{multirow,makecell}   
\usepackage{textcomp}
\usepackage{wasysym}
\usepackage{setspace}
\usepackage{verbatim}
\usepackage{float}
\usepackage{ulem}
\usepackage[utf8]{inputenc}

\usepackage[text={17.2cm,25.2cm},centering]{geometry}

\numberwithin{equation}{section}

\begin{document}
	\title{\textbf{Phase diagram of holographic thermal dense QCD matter with rotation}}
	\author{Yan-Qing Zhao$^{1}$\footnote{zhaoyanqing@mails.ccnu.edu.cn }, Song He$^{2,3}$\footnote{hesong@jlu.edu.cn }, Defu Hou$^{1}$\footnote{houdf@mail.ccnu.edu.cn }, Li Li$^{4,5,6}$\footnote{ liliphy@itp.ac.cn}, and Zhibin Li$^{7}$\footnote{ lizhibin@zzu.edu.cn}}
	\date{}
	
	\maketitle
	
	\vspace{-10mm}
	
	\begin{center}
		{\it
			$^{1}$ Institute of Particle Physics and Key Laboratory of Quark and Lepton Physics (MOS),
Central China Normal University, Wuhan 430079, China\\ \vspace{1mm}
			
			$^{2}$ Center for Theoretical Physics and College of Physics, Jilin University,
	Changchun 130012, China\\ \vspace{1mm}			
			$^{3}$ Max Planck Institute for Gravitational Physics (Albert Einstein Institute), Am Muhlenberg 1, 14476 Golm, Germany\\ \vspace{1mm}			
			$^{4}$ CAS Key Laboratory of Theoretical Physics, Institute of Theoretical Physics,
	Chinese Academy of Sciences, Beijing 100190, China\\ \vspace{1mm}			
			$^{5}$ School of Fundamental Physics and Mathematical Sciences,
	Hangzhou Institute for Advanced Study, University of Chinese Academy of Sciences, Hangzhou 310024, China\\ \vspace{1mm}			
			$^{6}$ Peng Huanwu Collaborative Center for Research and Education, Beihang University, Beijing 100191, China\\ \vspace{1mm}			$^{7}$ School of Physics and Microelectronics, Zhengzhou University, Zhengzhou 450001, China\\ \vspace{1mm}
		}
		\vspace{10mm}
	\end{center}

	\begin{abstract}
We study the rotation effects of the hot and dense QCD matter in a non-perturbative regime by the gauge/gravity duality. We use the gravitational model that is designated to match the state-of-the-art lattice data on the thermal properties of (2+1)-flavor QCD and predict the location of the critical endpoint and the first-order phase transition line at large baryon chemical potential without rotation. After introducing the angular velocity via a local Lorentz boost, we investigate the thermodynamic quantities for the system under rotation in a self-consistent way. We find that the critical temperature and baryon chemical potential associated with the QCD phase transition decrease as the angular velocity increases. Moreover, some interesting phenomena are observed near the critical endpoint. We then construct the 3-dimensional phase diagram of the QCD matter in terms of temperature, baryon chemical potential, and angular velocity. As a parallel investigation, we also consider the gravitational model of $SU(3)$ pure gluon system, for which the 2-dimensional phase diagram associated with temperature and angular velocity has been predicted. The corresponding thermodynamic quantities with rotation are investigated.

	\end{abstract}
	
	\baselineskip 18pt
	\thispagestyle{empty}
	\newpage
	
	\tableofcontents
		
\section{Introduction}\label{sec:00_intro}
Understanding the phase structure of QCD matter is an interesting and fundamental challenge of high-energy physics. In particular, exploring the transition of the quark-gluon system is one of the main goals of the heavy ion collision experiments at the Relativistic Heavy Ion Collider (RHIC) and the Large Hadron Collider (LHC). To provide significant assistance to the experiments, many non-perturbative approaches have been proposed to study the QCD phase diagram under various conditions, such as the Lattice QCD \cite{Braguta:2021ucr, Scherzer:2020kiu}, the Nambu-Jona Lasinio (NJL) model \cite{Wang:2018sur}, and the hadron resonance gas model \cite{Fujimoto:2021xix}. Due to the famous sign problem, the state-of-the-art lattice QCD simulation can only provide reliable physical information at zero chemical potential. Although one can extrapolate lattice data to small chemical potential, the relevant results are still lacking. An alternative non-perturbative approach is to apply the gauge/gravity duality \cite{Maldacena:1997re, Gubser:1998bc, Witten:1998qj, Witten:1998zw} to construct holographic QCD models to describe QCD matters in terms of top-down approaches\cite{Babington:2003vm, Kruczenski:2003uq, Sakai:2004cn, Sakai:2005yt}  and bottom-up approaches \cite{Gubser:2008yx, Gursoy:2008bu}. In particular, several attempts were made to apply holographic QCD models \cite{DeWolfe:2010he, DeWolfe:2011ts, Cai:2012xh, Alho:2013hsa, Grefa:2021qvt, He:2013qq, Chen:2018vty, He:2020fdi} to understand the QCD phase diagram.

Recently, the rotational effect of quark matter has attracted much attention. A large angular momentum is generated in decentralized heavy ion collisions \cite{Liang:2004ph, Huang:2011ru, Becattini:2007sr, Csernai:2013bqa, Jiang:2016woz, Deng:2016gyh}. Although most of the angular momentum is carried by the so-called ``spectators", a significant fraction is still carried by the so-called ``participants" \cite{Jiang:2016woz}. Star collaboration estimates the angular velocity of the quark-gluon-plasma (QGP) fireball to be $\omega\approx (9\pm 1)\times 10^{21} s^{-1}$ \cite{STAR:2017ckg}. Such a large angular velocity leads to interesting phenomena, such as the chiral vortical effect \cite{Kharzeev:2015znc, Rogachevsky:2010ys, Yang:2021hor}, the polarization effect \cite{Guo:2021udq}, and the effect of rotation on the chiral phase transition \cite{Chernodub:2016kxh, Wang:2018sur}, deconfinement phase transition \cite{Braguta:2021jgn, Braguta:2020biu, Chen:2020ath, Braga:2022yfe, Chernodub:2020qah, Ebihara:2016fwa} and other physical quantities \cite{Hou:2021own, Chen:2022obe, Zhou:2021sdy}.

The phase structure and thermodynamic properties of quark matter can change to an extreme degree upon rotation. The rotation effect on the chiral phase transition of quark matter has been studied in the literature using effective field theories. It has been shown that a nonzero angular velocity could suppress the chiral condensation in the NJL model~\cite{Wang:2018sur}. Moreover, there exists a nontrivial critical endpoint (CEP) in the $T-\omega$ plane for the chiral phase transition. A mixed inhomogeneous phase was discussed in \cite{Chernodub:2020qah}, and the authors claimed that the mixed phase separates the confinement and deconfinement regions. By applying the hadron resonance gas model \cite{Fujimoto:2021xix}, it was found that the deconfinement temperature decreases with rotation. In addition, the effect of inverse magneto-rotational catalysis was found when considering the phase structure of rotating hot and magnetized quark matter \cite{Sadooghi:2021upd}. In holography, the rotation effect has been considered in some cases. For example, the authors of \cite{Chen:2020ath} studied the phase transition of deconfinement under rotation in a pure gluon and 2-flavor systems using the potential reconstruction approach in holographic QCD literature \cite{Li:2011hp, Cai:2012xh, Li:2012ay}\footnote{One can also refer to more recent developments on this approach \cite{Arefeva:2018hyo, Ren:2019lgw, Bohra:2019ebj, Bohra:2020qom, He:2020fdi}.}. It was found that the phase transition in the $T-\omega$ plane is always a crossover for the 2-flavor system with a small chemical potential $\mu_B$. However, for a pure gluon system, the transition is first-order at vanishing chemical potential, and a CEP exists as chemical potential increases. {Moreover, using the Tolman-Ehrenfest formula adapted for rotation, the author of \cite{Chernodub:2020qah} suggested a kinematic relation between the phase transition temperature and the rotational velocity $\omega$, i.e. $T_c(\omega)/T_c(0)=\sqrt{1-\omega^2\ell^2}$.} Nevertheless, some recent lattice QCD simulations \cite{Braguta:2021ucr, Braguta:2021jgn} have yielded the opposite behavior for a pure gluon system. The deconfinement phase transition temperature increases with angular velocity $T_c(\omega)/T_c(0)=1+C_2\,\omega^2$ with  $C_2$ a positive constant. The such discrepancy might suggest the existence of other unknown physical relations.

In this work, we would like to apply the holographic approach to study the rotational effects in the $2+1$ flavor QCD matter and the pure $SU(3)$ gauge theory. The Einstein-Maxwell dilaton (EMD) model has been used in several papers \cite{Gubser:2008yx, DeWolfe:2010he, DeWolfe:2011ts} to study the QCD phase structure (see~\cite{Chen:2022goa} for a recent review). We will adopt the holographic models for the $2+1$ flavor QCD matter, and the pure $SU(3)$ gauge theory recently introduced in \cite{Cai:2022omk} and \cite{He:2022toapp}, respectively. Both models are fixed by quantitatively matching the state-of-the-art lattice data. Moreover, the $2+1$ flavor model provides a reasonable CEP in the $T-\mu_B$ plane, thus could be used to study the behavior of CEP under rotation. We will apply a Lorentz boost to mimic the rotating QCD matter. We also provide self-consistent thermodynamic relations under rotation and introduce the corresponding thermodynamic quantities, e.g., entropy, temperature, pressure, etc. Finally, we will investigate how rotation changes the QCD phase diagram.

The rest of this manuscript is organized as follows. In section \ref{sec:01_model}, we will introduce the holographic model and construct the rotating metric. In section \ref{sec:02_eos}, we will define the thermodynamic quantities using holographic renormalization. The computation of the Polyakov loop is provided in section~\ref{Ploop}. In section \ref{sec:03_result}, we present our numerical results of the rotation effect. We end with a summary and discussion in section \ref{sec:04summary}.

\section{Holographic QCD model}\label{sec:01_model}
\begin{figure}
  \centering
   \includegraphics[width=0.49\textwidth]{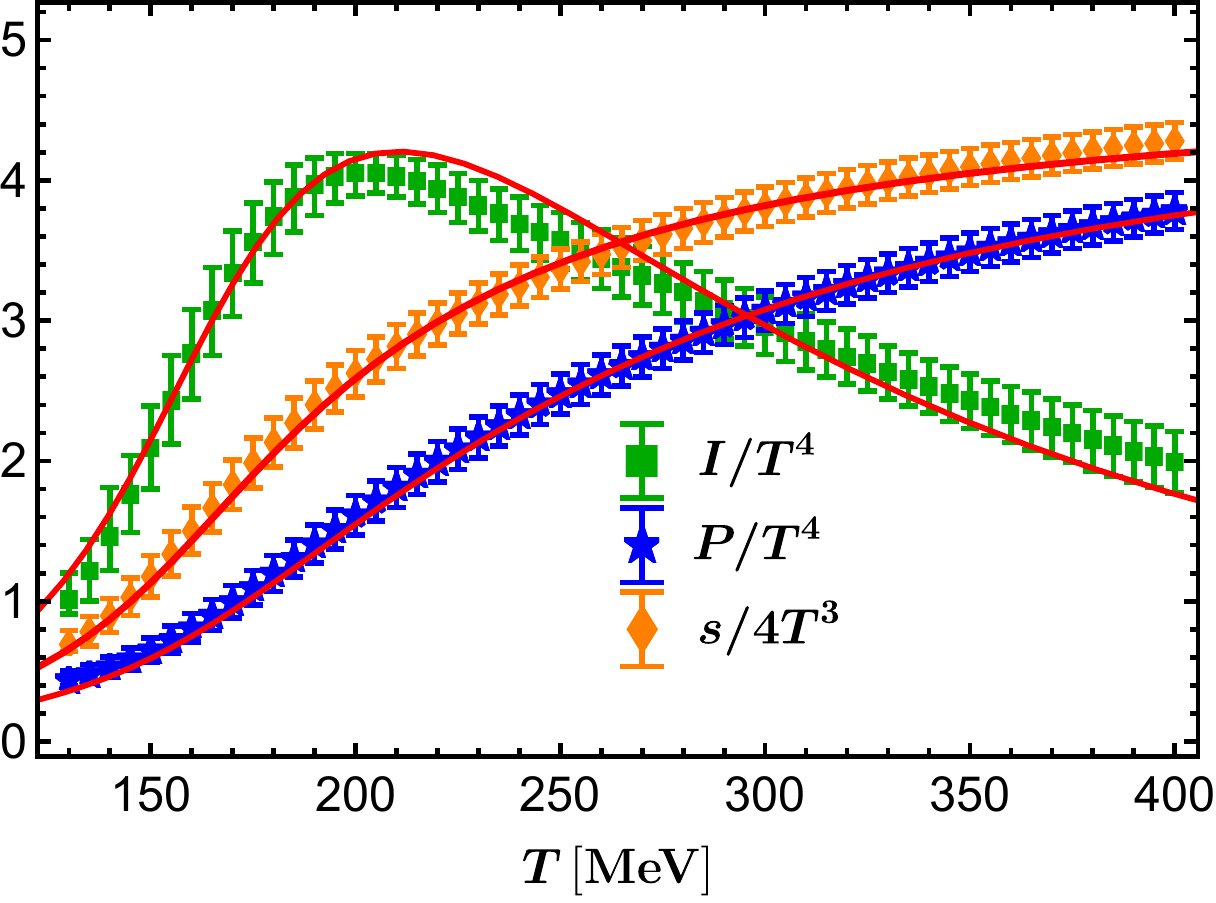}
  \includegraphics[width=0.49\textwidth]{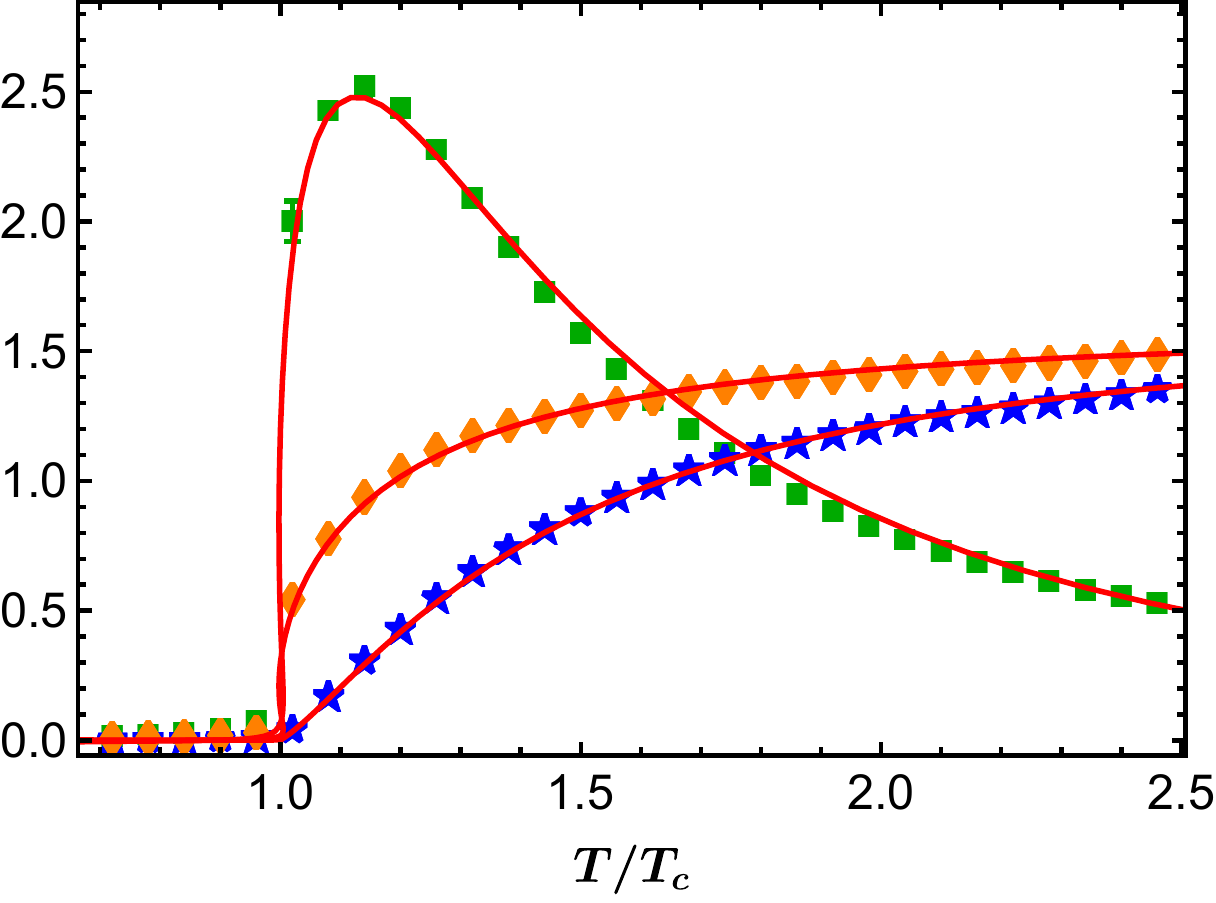}\\
  \caption{The temperature dependence of entropy density $s$, pressure $P$, and trace anomaly $I$ for the 2+1 flavor (left) and pure gluon (right) systems. Our holographic results(solid curves)~\cite{Cai:2022omk,He:2022toapp} quantitatively match the state-of-the-art lattice data (with error bar)~\cite{HotQCD:2014kol,Caselle:2018kap}.}\label{fig00}
\end{figure}

We set up a 5-dimensional holographic QCD model in EMD gravitational system. The action takes the following form \cite{Cai:2022omk,He:2022toapp}:
\begin{equation}\label{eq1}
  S_M =\frac{1}{2\kappa_N^2}\int d^5x\sqrt{-g}[\mathcal{R}-\frac{1}{2}\nabla_\mu\phi\nabla^\mu\phi-\frac{Z(\phi)}{4}F_{\mu\nu}F^{\mu\nu}-V(\phi)]\,,
\end{equation}
where the potential and kinetic functions read
\begin{eqnarray}
 &&  V(\phi)=-12 \cosh\left[c_1\phi\right]+\left(6c_1^2-\frac{3}{2}\right)\phi^2+c_2\phi^6, \\
&&Z(\phi)=\frac{1}{1+c_3}{\mathrm sech}[c_4\phi^3]+\frac{c_3}{1+c_3}e^{-c_5\phi}.
\end{eqnarray}
The action includes three fields: a metric field $g_{\mu\nu}$ that corresponds to spacetime geometry, a dilaton field $\phi$ encoding the running of the gauge coupling, and a vector field $A_\mu$ introducing a finite chemical potential and baryon density. Note that for pure $SU(3)$ gauge theory, there is no quark degree of freedom, so we close the gauge field $A_\mu$ by setting $Z(\phi)=0$. The parameters for pure gauge theory and $2+1$ flavor models could be fixed by fitting the lattice QCD data at zero net-baryon density \cite{Boyd:1996bx, Gupta:2007ax, HotQCD:2014kol, Borsanyi:2021sxv,Caselle:2018kap}. The results are shown in TABLE \ref{table1}. Here $\kappa_N^2$ is the effective Newton constant, and $\phi_s$ is the source of the dual scalar operator in the boundary field theory, which essentially breaks the conformal symmetry and fixes the energy scale of the system. The parameter $b$ is from the holographic renormalization and is necessary to satisfy the lattice QCD simulation at $\mu_B=0$. {We compare various thermodynamic quantities from our holographic setup~\cite{Cai:2022omk,He:2022toapp} with lattice simulation in Fig.~\ref{fig00}. One can see that the temperature dependence of all those quantities agree well with lattice results for the $2+1$ flavor~\cite{HotQCD:2014kol} and pure gluon~\cite{Caselle:2018kap} systems.}

\begin{table}[h!]
    \centering
    \setlength{\tabcolsep}{3.5mm}{
    \begin{tabular}{|c|c|c|c|c|c|c|c|c|}
     \cline{1-9}
    model  &  $c_1$  &  $c_2$  & $c_3$ & $c_4$  & $c_5$ & $\kappa_N^2$ & $\phi_s $(GeV) & b \\ \cline{1-9}
    pure $SU(3)$  & 0.735   &  0  &   &   &  & $2\pi(4.88)$ & 1.523 & -0.36458\\ \cline{1-9}
    2+1 flavor  &  0.710  &  0.0037  & 1.935  & 0.085  & 30 & $2\pi(1.68)$  &  1.085 & -0.27341 \\ \cline{1-9} 
    \end{tabular}}
    \caption{Parameters for the pure $SU(3)$ gauge theory and $2+1$ flavor models~\cite{Cai:2022omk,He:2022toapp} by matching the lattice simulation.}
    \label{table1}
\end{table}

We choose the background as
\begin{align}\label{eq2}
  ds^2 &=-e^{-\eta(r)}f(r)dt^2+\frac{dr^2}{f(r)}+r^2(d x_1^2+d x_2^2+d x_3^2),\nonumber\\
  \phi &=\phi(r),\quad\quad\quad A_t=A_t(r),
\end{align}
where $r$ is the holographic radial coordinate. Here $r\rightarrow \infty$ corresponds to the AdS boundary, and the horizon of the black hole is located at $r=r_h$ where $f(r_h)=0$. The Hawking temperature and entropy density can be calculated as
\begin{equation}\label{eq3}
  T=\frac{1}{4\pi}f'(r_h)e^{-\eta(r_h)/2},  \quad  s=\frac{2\pi}{\kappa_N^2}r_h^3.
\end{equation}
More details of the holographic model can be found in \cite{Cai:2022omk}. 

To introduce the rotation effect, we split the 3-dimensional space into two parts as $\mathcal{M}_3=\mathbb{R}\times \Sigma_2$. Then the metric becomes to
\begin{equation}\label{eq4}
  ds^2=-f(r)e^{-\eta(r)}dt^2+\frac{dr^2}{f(r)}+r^2\ell^2d\theta^2+r^2 d \sigma^2\,.
\end{equation}
where $d \sigma^2$ denotes the line element of $\Sigma_2$. We assume the system that has an angular velocity $\omega$ with a fixed radius $\ell$, and consider the following local Lorentz boost~\cite{Erices:2017izj, BravoGaete:2017dso, Sheykhi:2010pya, Awad:2002cz, Nadi:2019bqu, Chen:2020ath}
\begin{equation}\label{eq5}
  t\rightarrow \frac{1}{\sqrt{1-\omega ^2 \ell ^2}}(\hat{t}+\omega\ell^2\hat{\theta}), \quad \theta\rightarrow \frac{1}{\sqrt{1-\omega ^2 \ell ^2}}(\hat{\theta}+\omega \hat{t})\,.
\end{equation}
Since one can not violate the bound of the light velocity, the metric is well-defined for $\omega ^2 \ell ^2 < 1$.  The corresponding metric can be written as
\begin{equation}\label{eq6}
  d \hat{s}^2=g_{\mu\nu}d\hat{x}^\mu d\hat{x}^\nu=-N(r)d \hat{t}^2+\frac{d r^2}{f(r)}+R(r)(d\hat{\theta}+Q(r)d \hat{t})^2+r^2 d \sigma^2\,,
\end{equation}
where
\begin{align}\label{eq7}
  N(r) &= \frac{r^2  f(r) \left(1-\omega ^2 \ell^2\right)}{r^2  e^{\eta (r)}-\omega ^2 \ell ^2 f(r)}\,,\nonumber \\
  R(r) &=\frac{r^2 \ell ^2-\omega ^2 \ell ^4 f(r) e^{-\eta(r)}}{1-\omega ^2 \ell ^2}\,, \\
  Q(r) &=\frac{\omega  \left(f(r)-r^2 e^{\eta(r)}\right)}{\omega ^2 \ell ^2 f(r)-r^2 e^{\eta(r)}}\,.\nonumber
\end{align}
This dual gravity background provides a first approximation to rotating nuclear matter. The null generator of the horizon is $\xi=\partial_{\hat{t}}-\omega\partial_{\hat{\theta}}$ with $\omega$ the angular velocity. Then the baryon chemical potential is given by $\hat{\mu}_B=A_\mu\xi^\mu|_{r=\infty}-A_\mu\xi^\mu|_{r=r_h}$.

The resulting Hawking temperature and entropy density are given by
\begin{equation}\label{eq9}
  \hat{T}=T\sqrt{1-\omega^2\ell^2}=\frac{1}{4\pi}f'(r_h)e^{-\eta(r_h)/2}\sqrt{1-\omega^2\ell^2},  \quad  \hat{s}=\frac{s}{\sqrt{1-\omega ^2 \ell ^2}}=\frac{2\pi}{\kappa_N^2}r_h^3\frac{1}{\sqrt{1-\omega ^2 \ell ^2}}\,, 
\end{equation}
where $T$ and $s$ are, respectively, the Hawking temperature and entropy density in the static frame~\eqref{eq3}.

\section{Thermodynamics}\label{sec:02_eos}
In this section, we will analyze the related thermodynamic quantities. Firstly, the free energy density of the system can be defined as
\begin{equation}
   \hat{ \Omega}=-\frac{\hat{T}}{\hat{V}}(S_M+\hat{S}_{\partial})_{on-shell}\,,
\end{equation}
with $S_M$ the action in the bulk and $\hat{S}_{\partial}$ the boundary term. In this work, we choose the same form of boundary term as in \cite{Cai:2022omk}
\begin{equation}\label{eq10}
  \hat{S}_{\partial}=\frac{1}{2\kappa_N^2}\int_{r\rightarrow \infty}d^4x\sqrt{-\hat{h}}\bigg[2\hat{K}-6-\frac{1}{2}\phi^2-\frac{6c_1^4-1}{12}\phi^4\ln[r]-b\phi^4+\frac{1}{4}F_{\rho\lambda}F^{\rho\lambda}\ln[r]\bigg]\,,
\end{equation}
where $\hat{h}_{\mu\nu}$ is the induced metric at the boundary, $\hat{K}_{\mu\nu}$ is the extrinsic curvature defined by the outward pointing normal vector to the boundary, and  $\hat{K}$ is its trace. Then it can be derived directly from the equations of motion that~\footnote{Note that although the metric \eqref{eq6} takes a different form from the one in the static frame \eqref{eq4}, the equations of motion remain unchanged. This is also why we can fix the boundary term in the same form within the static frame, as shown in \eqref{eq10}.}
\begin{equation}\label{eqfree}
 \hat{\Omega}=-\frac{1}{2\kappa_N^2}\big(-f_\upsilon+\phi_s\phi_\upsilon+\frac{3-48b-8c_1^4}{48}\phi_s^4\big)\,,
\end{equation}
with $\phi_s$, $\phi_v$ and $f_v$ the UV expansion coefficients of $\phi(r)$ and $f(r)$.
\begin{equation}
 \phi(r)=\frac{\phi_s}{r}+\frac{\phi_v}{r^3}+\cdots,\quad f(r)=r^2+\cdots+\frac{f_v}{r^2}+\cdots\,.
\end{equation}

The baryon density $\hat{\rho}_B$ and the angular momentum $J$ can be obtained from the relation
\begin{equation}
   d \hat{\Omega}=-\hat{s} d\hat{T}-\hat{\rho}_B d \hat{\mu}_B-J d\omega\,,
\end{equation}
with $\hat{T}$ and $\hat{s}$ given in \eqref{eq9}. One can easily check that $\partial\hat{\Omega}/\partial \hat{T}=-\hat{s}$ for fixed $\hat{\mu}_B$ and $\omega$. The gauge field $\hat{A}_\mu$ reads
\begin{equation}\label{eq25}
  \hat{A}_\mu=\frac{1}{\sqrt{1-\omega^2\ell^2}}A_t\delta^t_\mu+\frac{\omega \ell^2}{\sqrt{1-\omega^2\ell^2}}A_t\delta^\theta_\mu\,.
\end{equation}
Then the baryon chemical potential can be written as
\begin{equation}\label{eq27}
  \hat{\mu}_B=\mu_B\sqrt{1-\omega^2\ell^2}\,,
\end{equation}
with $\mu_B$ the chemical potential in the rest frame. Moreover, the baryon charge density is given by $\hat{\rho}_B=\rho_B/\sqrt{1-\omega ^2 \ell ^2}$. The energy-momentum tensor of the dual field theory can be calculated as
\begin{align}\label{eq11}
  \hat{T}_{\mu\nu} &= \lim_{r\to \infty}\frac{2}{\sqrt{-\hat{g}}}\frac{\delta(\hat{S}+\hat{S}_{\partial})_{on-shell}}{\delta \hat{g}^{\mu\nu}}\nonumber\\
             &=\frac{1}{2\kappa_N^2}\lim_{r\to \infty}r^2\bigg[2(\hat{K} \hat{h}_{\mu\nu}-\hat{K}_{\mu\nu}-3\hat{h}_{\mu\nu})-(\frac{1}{2}\phi^2+\frac{6c_1^4-1}{12}\phi^4\ln[r]+b\phi^4)\hat{h}_{\mu\nu}\nonumber\\
             & -(F_{\mu\rho}F_{\nu}^\rho-\frac{1}{4}\hat{h}_{\mu\nu}F_{\rho\lambda}F^{\rho\lambda})\ln[r]\bigg]\,.
\end{align}
We choose the coordinates on the boundary as $x^{\mu}=(t,~\ell\theta,~x,~y)$ where $(x,~y)$ denotes the coordinates on the surface $\Sigma_2$ and $\ell\theta$ corresponds to the direction of the local Lorentz boost. Then the non-vanishing components are given by
\begin{align}\label{eq20}
  \hat{\epsilon} &:=\hat{T}_{00}=\frac{ \epsilon+\omega^2\ell^2 P}{1-\omega^2 \ell^2}\,,\notag \\
  \hat{P}_1 &:=\hat{T}_{11}=\frac{1}{\ell^2}\hat{T}_{\theta\theta}=\frac{ P+\omega^2\ell^2 \epsilon}{1-\omega^2 \ell^2}\,,   \\
 \hat{ P}_2 &:=\hat{T}_{22}=P\,,   \notag \\
 \hat{ P}_3 &:=\hat{T}_{33}=P\,, \notag 
\end{align}
and
\begin{align}
    \hat{T}_{\theta t}=\frac{\omega \ell^2 (\epsilon+P)}{1-\omega^2\ell^2}\,,
\end{align}
where $\epsilon$ and $P$ are the energy density and pressure in the static frame (see \cite{Li:2020spf, Cai:2022omk} for more details). 

The angular momentum is related to $\hat{T}_{\theta t}$, i.e.
\begin{equation}\label{eqJ}
J=\frac{\omega \ell^2 (\epsilon+P)}{1-\omega^2\ell^2}\,.    
\end{equation}
As a double check, one has the following thermodynamic relation
\begin{equation}\label{eqrelation}
\hat{\Omega}=\hat{\epsilon}-\hat{T}\hat{s}-\hat{\mu}_B\hat{\rho}_B-\omega J=-P\,. 
\end{equation}
Moreover, the trace anomaly reads
\begin{equation}\label{eq22}
    \hat{I}=-\hat{T}^{\mu}_{\mu}=-T^{\mu}_{\mu}=I=\epsilon-3P\,,
\end{equation}
which is the same as the case without rotation. 

Besides the equation of state (EOS), the transport coefficients can also provide useful information for the QCD matter. The squared speed of sound can be calculated as~\footnote{In the absence of chemical potential and rotation, one has a simple relation $\hat{c}_s^2=\partial \ln{\hat{T}}/\partial \ln{\hat{s}}$ from~\eqref{eqcs}. We should stress that this relation is not valid at finite $\hat{\mu}_B$ and $\omega$.}
\begin{equation}\label{eqcs}
    \hat{c}_s^2=\frac{d P}{d\hat{\epsilon}}\Big|_{\hat{\mu}_B,\omega}\,,
\end{equation}
The specific heat and the second-order baryon susceptibility are given by
\begin{equation}
\hat{C}_v=\frac{d \hat{\epsilon}}{d\hat{T}}\Big|_{\hat{\mu}_B,\omega},\quad \hat{\chi}_2^B={{\frac{1}{\hat{T}^2}}}\frac{\partial\, \hat{\rho}_B}{\partial\,{\hat{\mu}_B}}\Big|_{\hat{T},\omega}=-{{\frac{1}{\hat{T}^2}}}\frac{\partial^2\, \hat{\Omega}}{\partial\,{\hat{\mu}_B}^2}\Big|_{\hat{T},\omega}\,, 
\end{equation}
respectively.

\section{Polyakov loop}\label{Ploop}
A good order parameter for the confinement phase transition in pure gauge theory is the Polyakov loop operator  $\langle \hat{\mathcal{P}} \rangle$, which is finite in the deconfined phase and becomes vanishing in the confined phase for pure gluon. Although it is not a suitable order parameter in the presence of massive quarks,  the behavior of $\langle \hat{\mathcal{P}} \rangle$ could also characterize some features of QCD matter.
Therefore, we will also study the rotating effect on the Polyakov loop.

With the geometric background \eqref{eq6}, one can calculate the expectation value of the Polyakov loop operator directly according to the holographic dictionary \cite{Andreev:2009zk}. It is convenient to redefine the holographic coordinate $r$ to be $z=1/r$ so that $z=0$ and $z=z_h$ correspond to the boundary and horizon of bulk spacetime, respectively. We also take the redefinition of fields: $f(r)=F(z)/z^2$, $\eta(r)=\Sigma(z)$, $\phi(r)=\Phi(z)$. Then the metric in the string frame becomes
\begin{align}
    ds_{string}^2 &=e^{\sqrt{\frac{2}{3}}\Phi(z)}d\hat{s}^2 \notag\\
    &=e^{\sqrt{\frac{2}{3}}\Phi(z)}(-\tilde{N}(z)d \hat{t}^2+\frac{d z^2}{z^2 F(z)}+\tilde{R}(z)(d\hat{\theta}+\tilde{Q}(z)d \hat{t})^2+z^{-2}d \sigma^2)\,,
\end{align}
with
\begin{align}
    & \tilde{N}(z)=\frac{F(z)(1-\omega^2\ell^2)}{z^2(e^{\Sigma(z)}-\omega^2\ell^2F(z))}\, ,\notag\\
    & \tilde{R}(z)=\frac{\ell^2-e^{-\Sigma(z)}\ell^4\omega^2 F(z)}{z^2(1-\omega^2\ell^2)}\, ,\\
    & \tilde{Q}(z)=\frac{\omega(e^{\Sigma(z)}- F(z))}{e^{\Sigma(z)}-\omega^2\ell^2F(z)}\, .\notag
\end{align}
And the Nambu-Goto action of the classic string shows
\begin{equation}
    S_{NG}=-\frac{1}{2\pi\alpha^\prime} \int d^2\xi \sqrt{ {\rm det} [g_{MN}(\partial_a X^M)(\partial_bX^N)]}\,,
\end{equation}
where $\alpha^\prime$ is the effective string tension. We choose the boundary of the string at $x_3=\pm\lambda/2$. Then the reduced metric on the string world sheet reads
\begin{equation}
    g_{ab}=e^{\sqrt{\frac{2}{3}} \Phi(z)}\left(\begin{matrix}
    \tilde{P}(z)^2 \tilde{R}(z)-\tilde{N}(z) & 0\\
   0 & \frac{z'(x_3)^2}{z^2 F(z)}+\frac{1}{z^2}
    \end{matrix}
    \right)\,,
\end{equation}
with the boundary conditions of $z(x_3)$ 
\begin{equation}
     z(\pm\frac{\lambda}{2})=0 \quad \text{or equivalently} \quad z(0)=z_0 \text{,} \quad  z'(0)={0}\, .
\end{equation}

The on-shell renormalized free energy for the Polyakov loop operator is given by
\begin{equation}
    \hat{F}_{ren}=\frac{1}{\pi\alpha' z_0}\left[-1+\int_0^1\frac{dv}{v^2}\left(\frac{e^{-\frac{1}{2}\Sigma(v z_0)+\sqrt{\frac{2}{3}}\Phi(v z_0)}\sqrt{F(v z_0)-e^{\Sigma(v z_0)}\ell^2\omega^2}}{\sqrt{F(v z_0)}\sqrt{1-\omega^2\ell^2}\sqrt{1-\tau(v)}} -1\right) \right]
\end{equation}
with $v=z/z_0$ and
\begin{equation}
    \tau(v)=v^4 e^{-\Sigma(z_0)+\Sigma(v z_0)+2\sqrt{\frac{2}{3}}\left( \Phi(z_0)-\Phi(v z_0 \right)}\frac{e^{\Sigma(z_0)}\omega^2 \ell^2-F(z_0)}{e^{\Sigma(v z_0)}\omega^2\ell^2-F(v z_0)}\,.
\end{equation}
Then one obtains the Polyakov loop operator
\begin{equation}
    \langle \hat{\mathcal{P}} \rangle=e^{-\frac{\hat{F}_{ren}|L\rightarrow \infty}{2\hat{T}}},
\end{equation}
where $L=\int_{-\frac{{\lambda}}{2}}^{\frac{{\lambda}}{2}} d{x_3} = 2\int_0^{z_0}\frac{dz}{z^\prime({x_3})}$  is the inter-quark distance.

\section{Rotation effects on QCD matter}\label{sec:03_result}
In this section, we will study how the rotation effects are presented in detail in the thermodynamic quantities and QCD phase diagram. We shall first focus on the 2+1 flavor model for which, as $\mu_B$ increases, the crossover is sharpened into a first-order line at the CEP located at ($T_c=105\text{MeV}$, $\mu_B=555\text{MeV}$) in the $T-\mu_B$ phase diagram without rotation \cite{Cai:2022omk}. Then, we will show the case for the pure gluon system.

\subsection{Thermodynamic quantities}

In this section, we would like to show the behaviors of different thermodynamic quantities, which are helpful in understanding the phase structure of QCD matter. 

\begin{figure}[h!]
  \centering
  \includegraphics[width=0.49\textwidth]{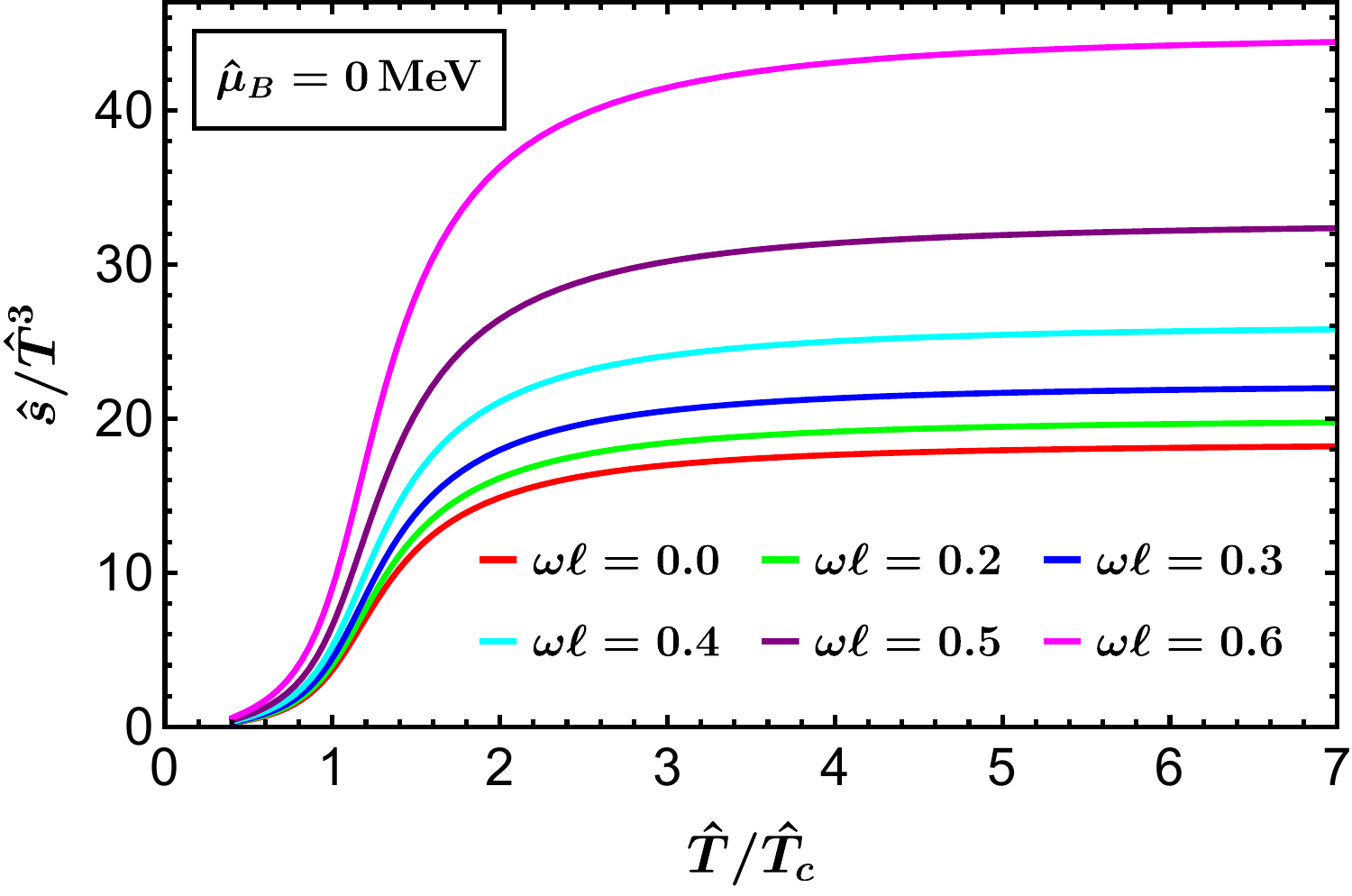}
  \includegraphics[width=0.49\textwidth]{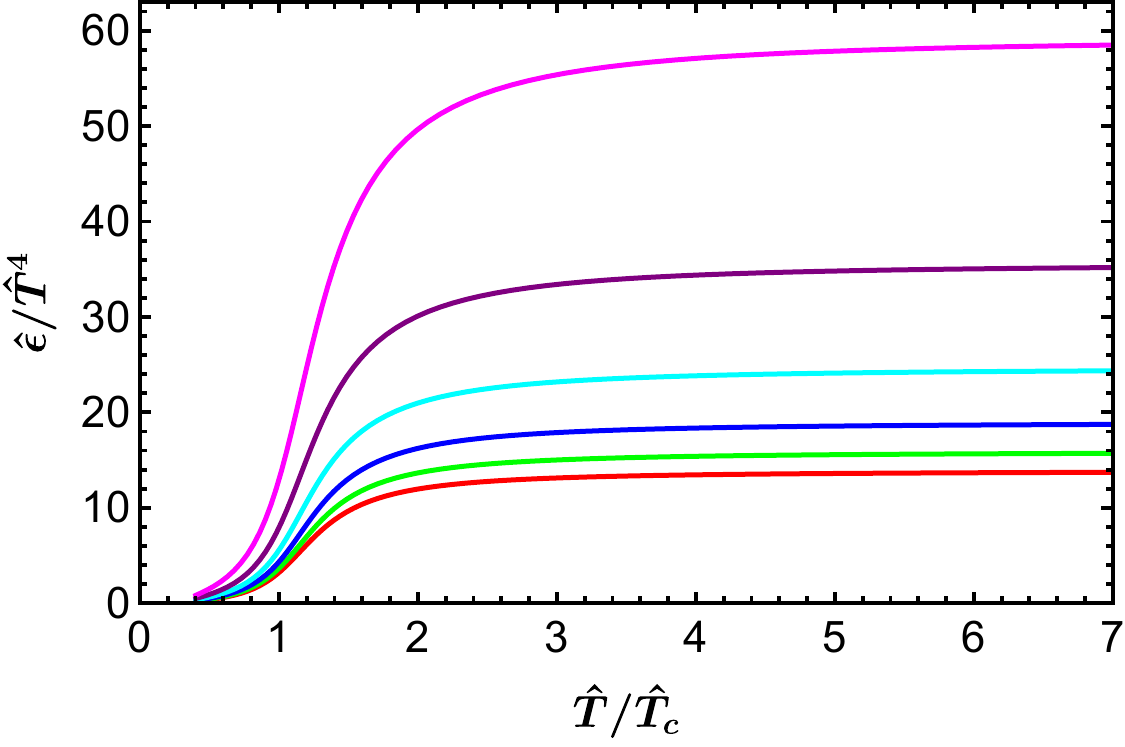}\\ 
  \includegraphics[width=0.49\textwidth]{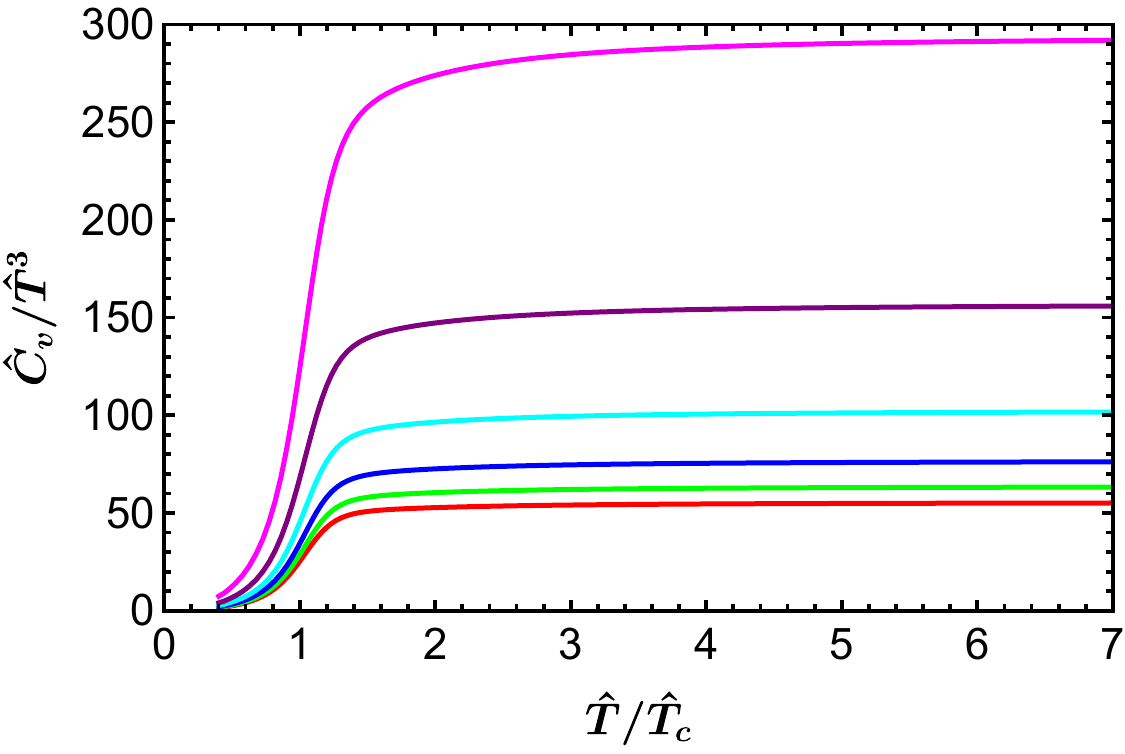}
  \includegraphics[width=0.49\textwidth]{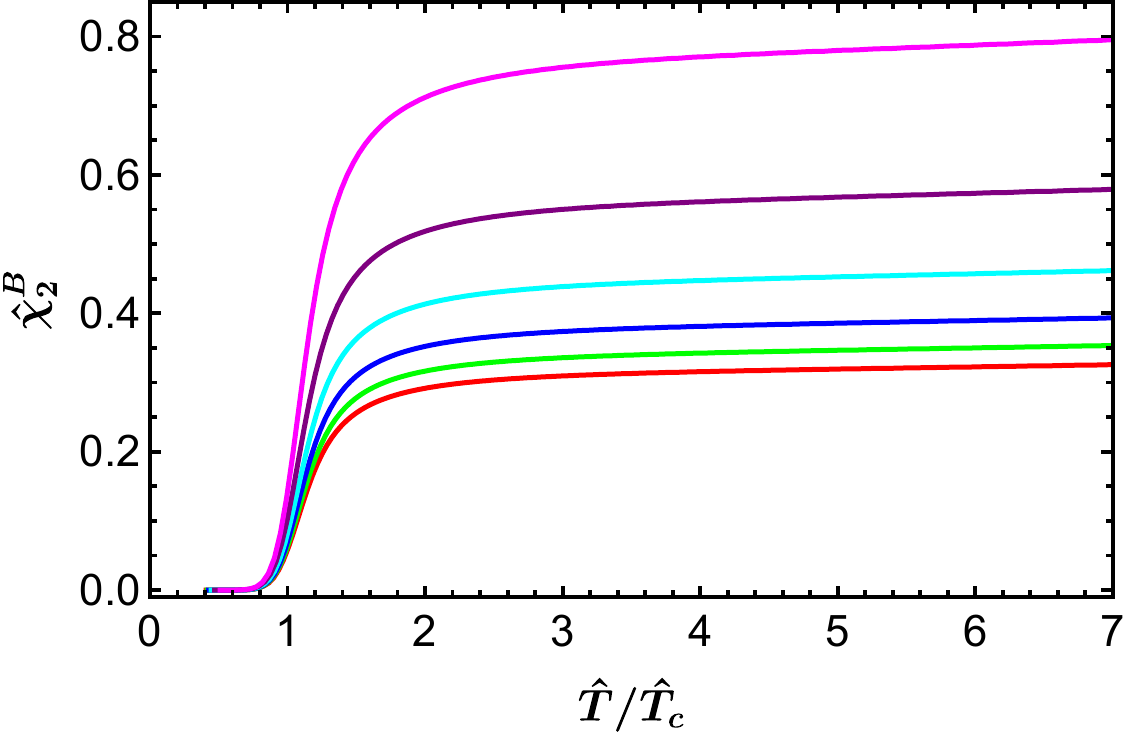}\\
  \caption{Entropy density $\hat{s}$, energy density $\hat{\epsilon}$, specific heat $\hat{C}_v$ and baryon susceptibility $\hat{\chi}_2^B$ as a function of normalized temperature $\hat{T}/\hat{T}_c$ at $\hat{\mu}_B=0$ for 2+1 flavor model. These quantities are all enhanced by increasing the angular velocity.}\label{fig4}
\end{figure}

The EOS with respect to the rotation velocity is presented in Fig.~\ref{fig4} at vanishing chemical potential. One finds that both the entropy density ${\hat{s}}$ and the energy density ${\hat{\epsilon}}$ increase as the angular velocity is increased and tend to be constants at the high-temperature limit. We also show the specific heat $\hat{C}_v$ and the baryon susceptibility $\hat{\chi}_2^B$ for different $\omega$. One finds that both are increased by increasing the angular velocity. In particular, all these quantities are dramatically enhanced by rotation. For an example, the energy density $\hat{\epsilon}/\hat{T}^4$ is enhanced almost $490\%$ from  $\omega\ell=0$ to $\omega\ell=0.6$ in the large $T$ limit.

\begin{figure}[h!]
  \centering
  \includegraphics[width=0.49\textwidth]{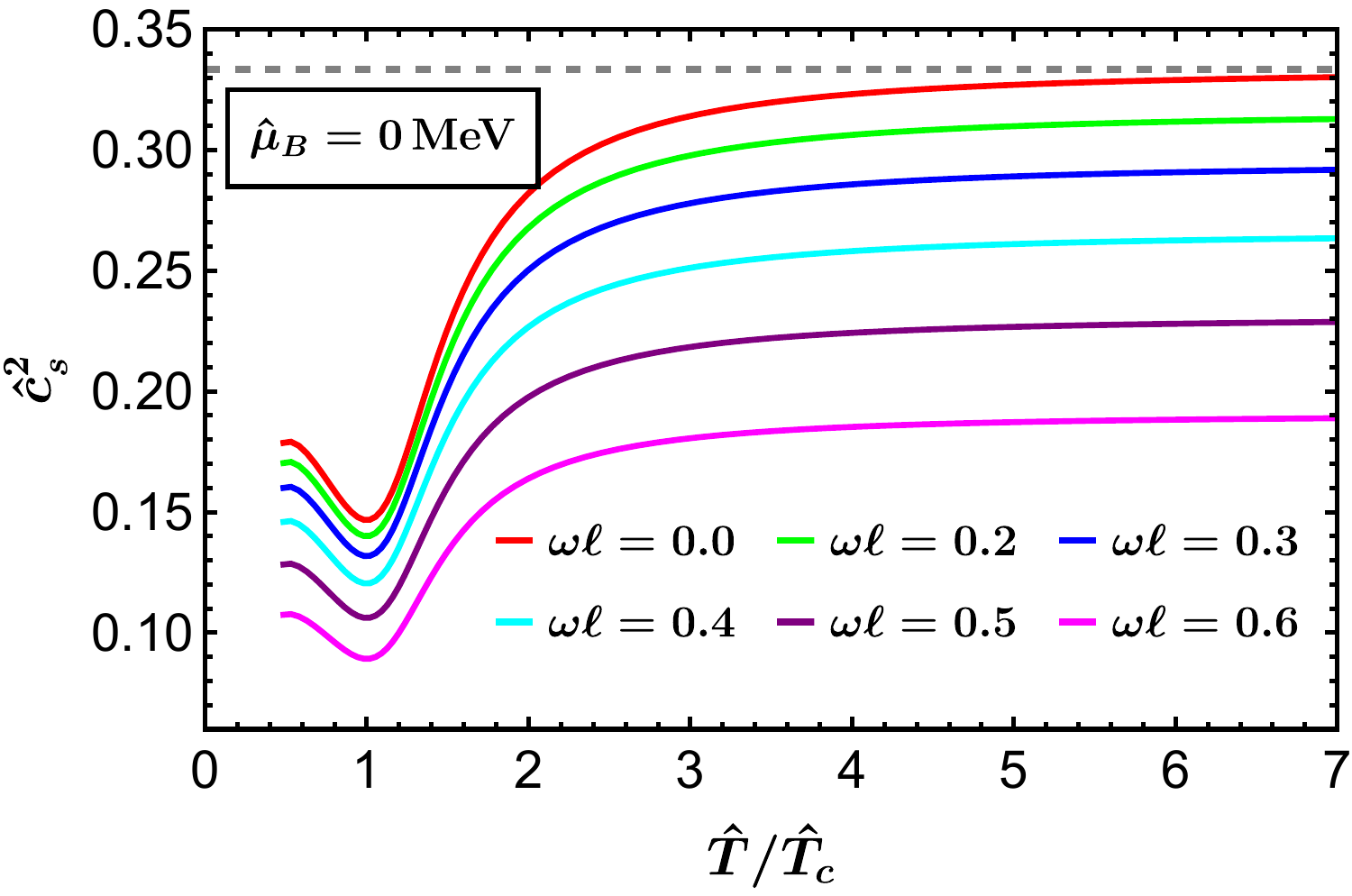}
  \includegraphics[width=0.49\textwidth]{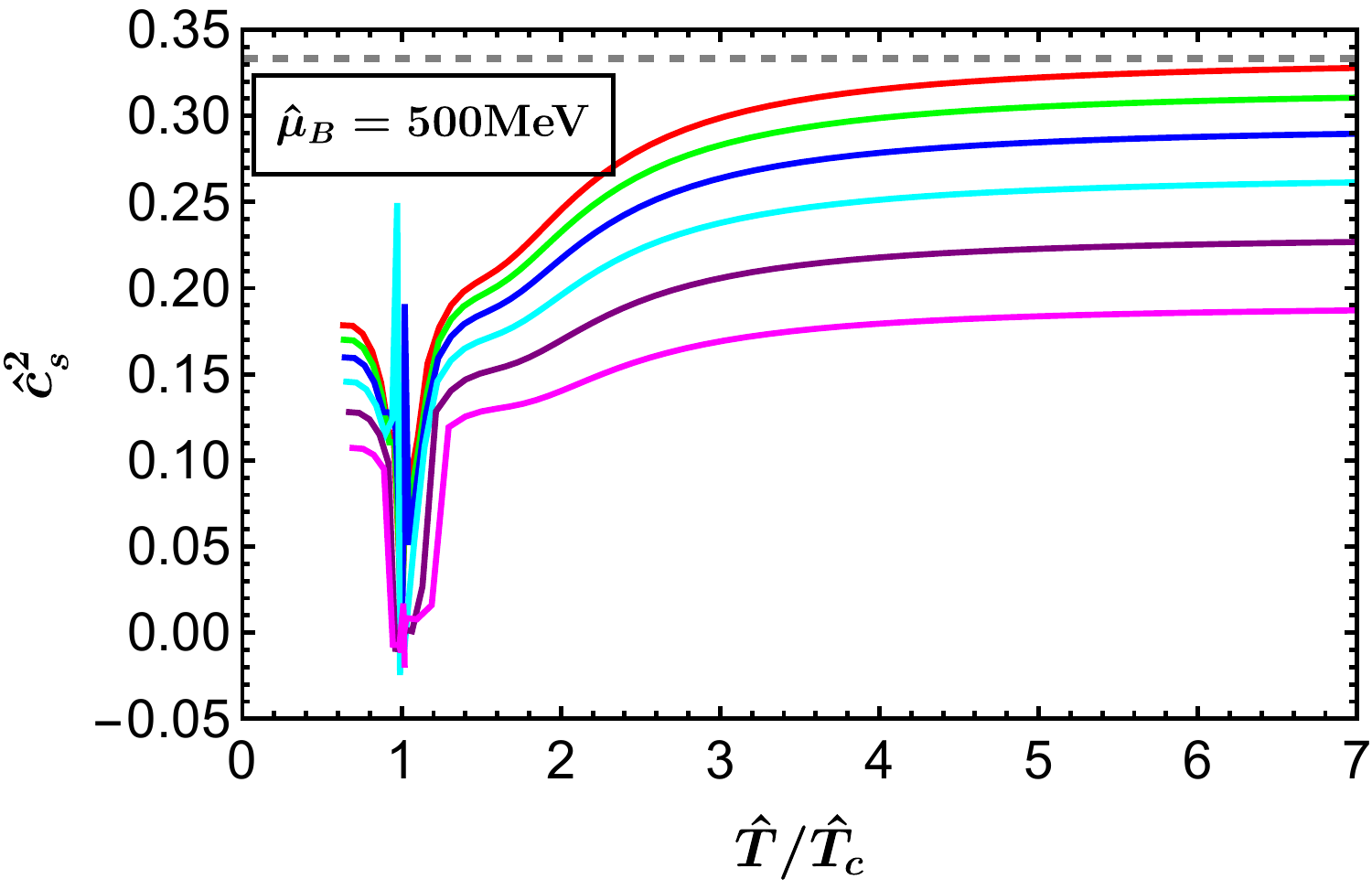}\\
  \caption{The temperature dependence of $\hat{c}_s^2$ at different angular velocity $\omega$. At $\hat{\mu}_B=0\,\text{MeV}$ (left) there is always crossover, while at $\hat{\mu}_B=500\,\text{MeV}$ (right) the  phase transition changes from crossover to first-order by increasing $\omega$.}\label{fig5p}
\end{figure}

Fig.~\ref{fig5p} shows the behavior of squared speed of sound $\hat{c}_s^2$ at $\hat{\mu}_B=0\,\text{MeV}$ (left panel) and $\hat{\mu}_B=500\,\text{MeV}$ (right panel) for different $\omega$. At vanishing chemical potential, there is a smooth crossover for which a good probe characterizing the drastic change of degrees of freedom between the QGP and the hadron resonances gas is the minimum speed of sound. Thus we define the pseudo-transition temperature $\hat{T}_c$ as the minimum of $\hat{c}_s^2$ in this case. It is clear that the speed of sound will approach a constant at a large temperature but will be suppressed by increasing the angular velocity, in sharp contrast to the enhancement in Fig.~\ref{fig4}. A similar feature can be found in the right panel at $\hat{\mu}_B=500\text{MeV}$. The new feature is that the smooth crossover becomes a sudden change when $\omega\ell>0.44$, signaling a first-order transition. 

\begin{figure}
  \centering
   \includegraphics[width=0.49\textwidth]{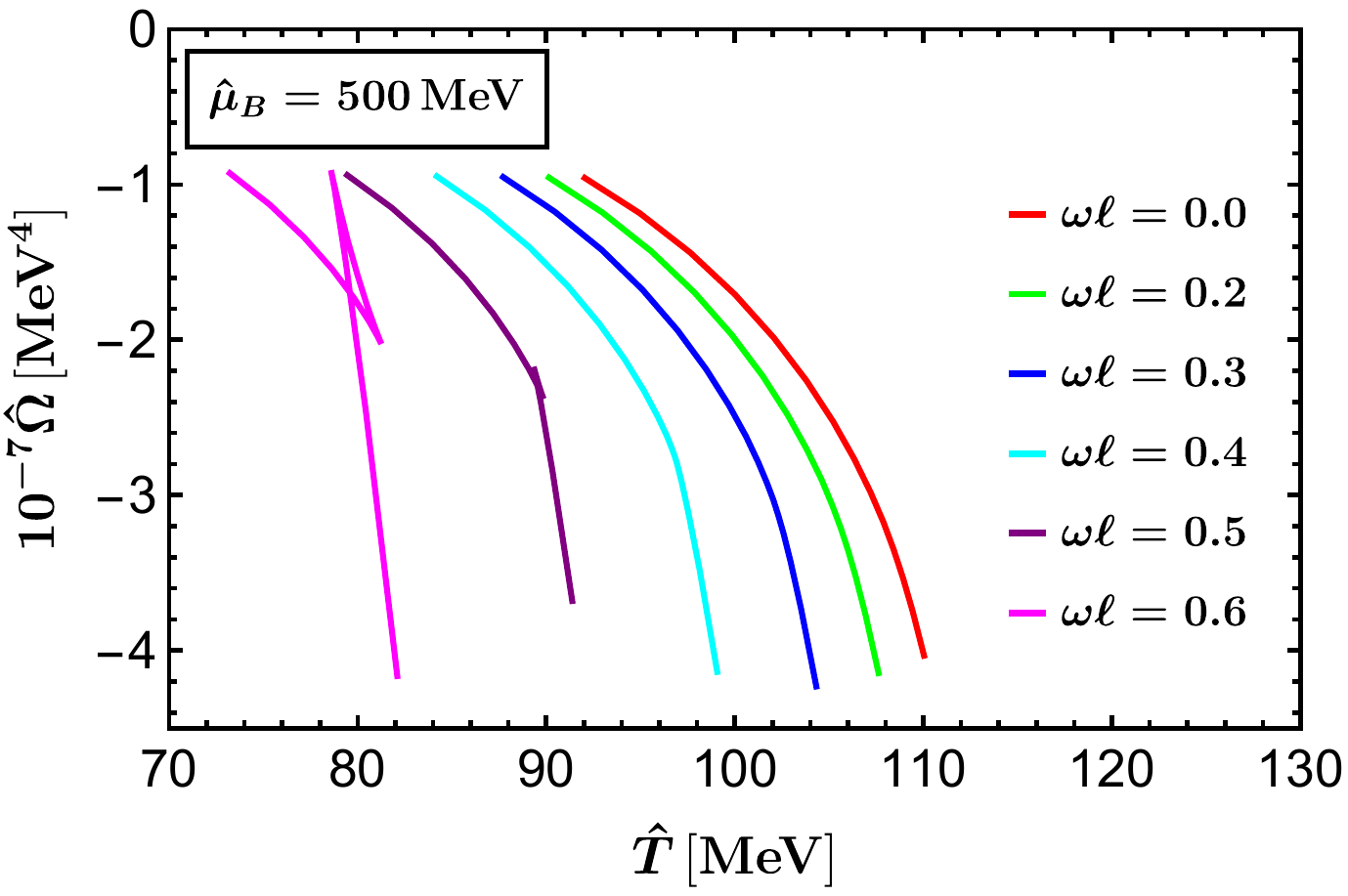}
  \includegraphics[width=0.49\textwidth]{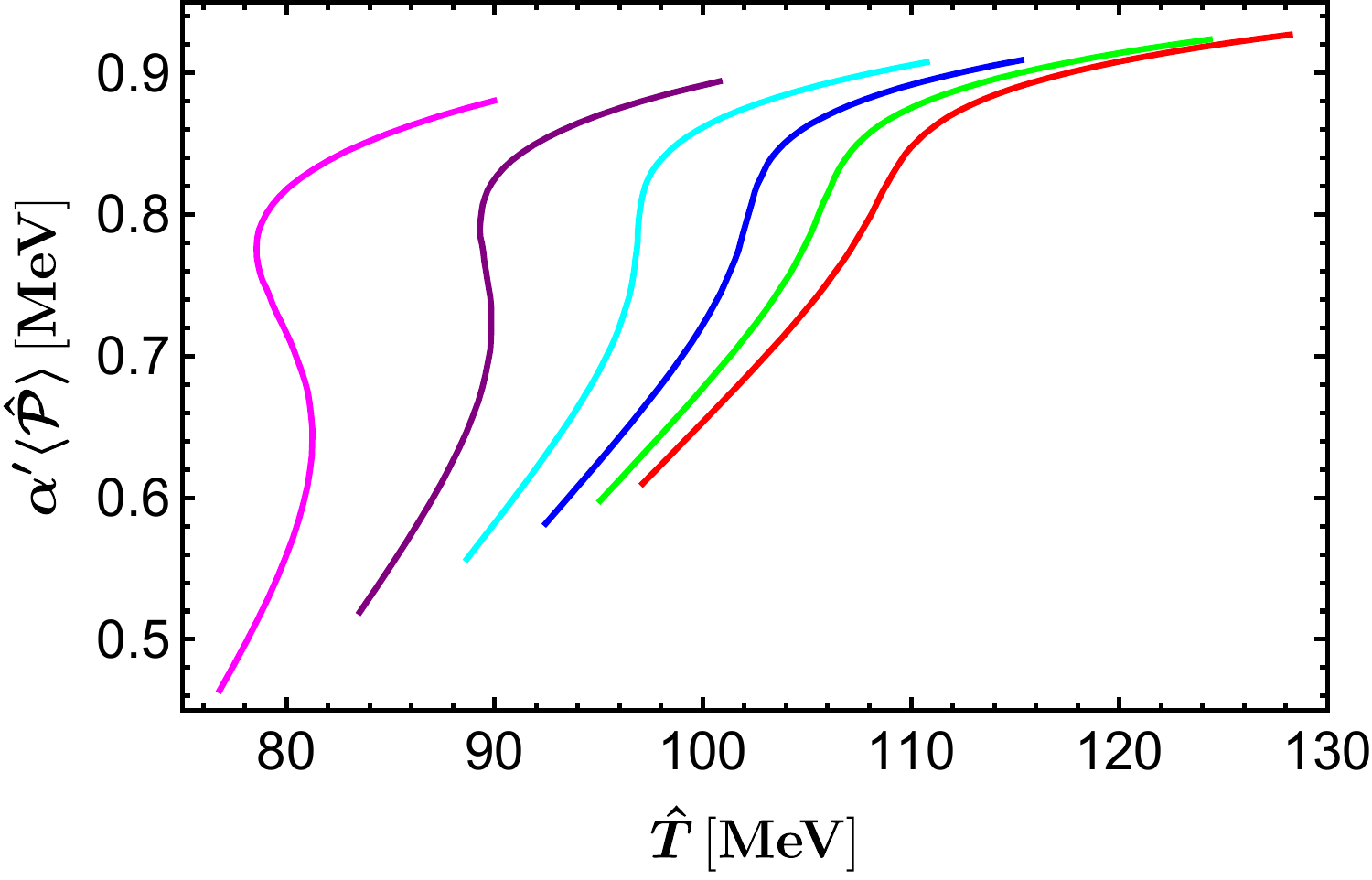}\\
  \caption{The free energy $\hat{\Omega}$ (left) and the Polyakov loop at different $\omega$ for $\hat{\mu}_B=500\text{MeV}$. The phase transition becomes first-order when $\omega\ell>0.44$.}\label{fig6}
\end{figure}

The temperature dependence of the free energy $\hat{\Omega}$ for different $\omega$ with $\hat{\mu}_B= 500\, \text{MeV}$ is presented in the left panel of Fig.~\ref{fig6}. While $\hat{\Omega}$ in the function of $\hat{T}$ decreases smoothly for $\omega\ell<0.44$, it becomes a swallowtail for $\omega\ell>0.44$, yielding a first-order phase transition. The corresponding behavior of Polyakov loop  $\langle \hat{\mathcal{P}} \rangle$ is given in the right panel of Fig.~\ref{fig6}. The Polyakov loop is not a good order parameter for the 2+1 flavor QCD system since the quark degrees of freedom break the $Z(N_c)$ symmetry. Nevertheless, it could be an effective order parameter. One finds that $\langle \hat{\mathcal{P}} \rangle$ is non-vanishing in the low-temperature phase. It increases rapidly near the pseudo-transition temperature in the crossover region when $\omega\ell<0.44$  and becomes multi-valued in the first-order transition region when $\omega\ell>0.44$. Thus, one obtains that the increase of angular velocity will trigger a first-order phase transition, consistent with the sudden change of $\hat{c}_s^2$ in Fig.~\ref{fig5p}.

\begin{figure}
  \centering
  \includegraphics[width=0.49\textwidth]{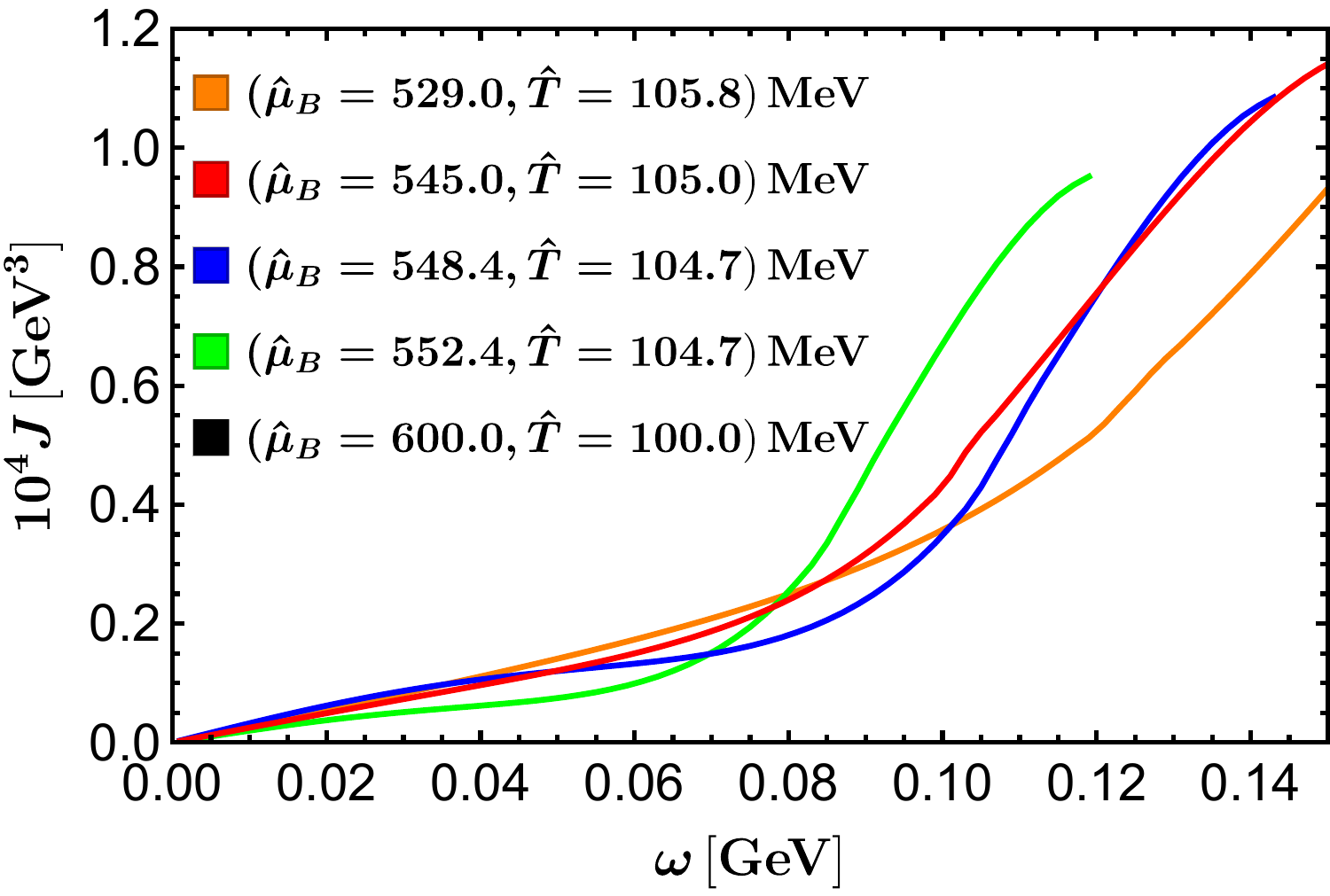}
  \includegraphics[width=0.49\textwidth]{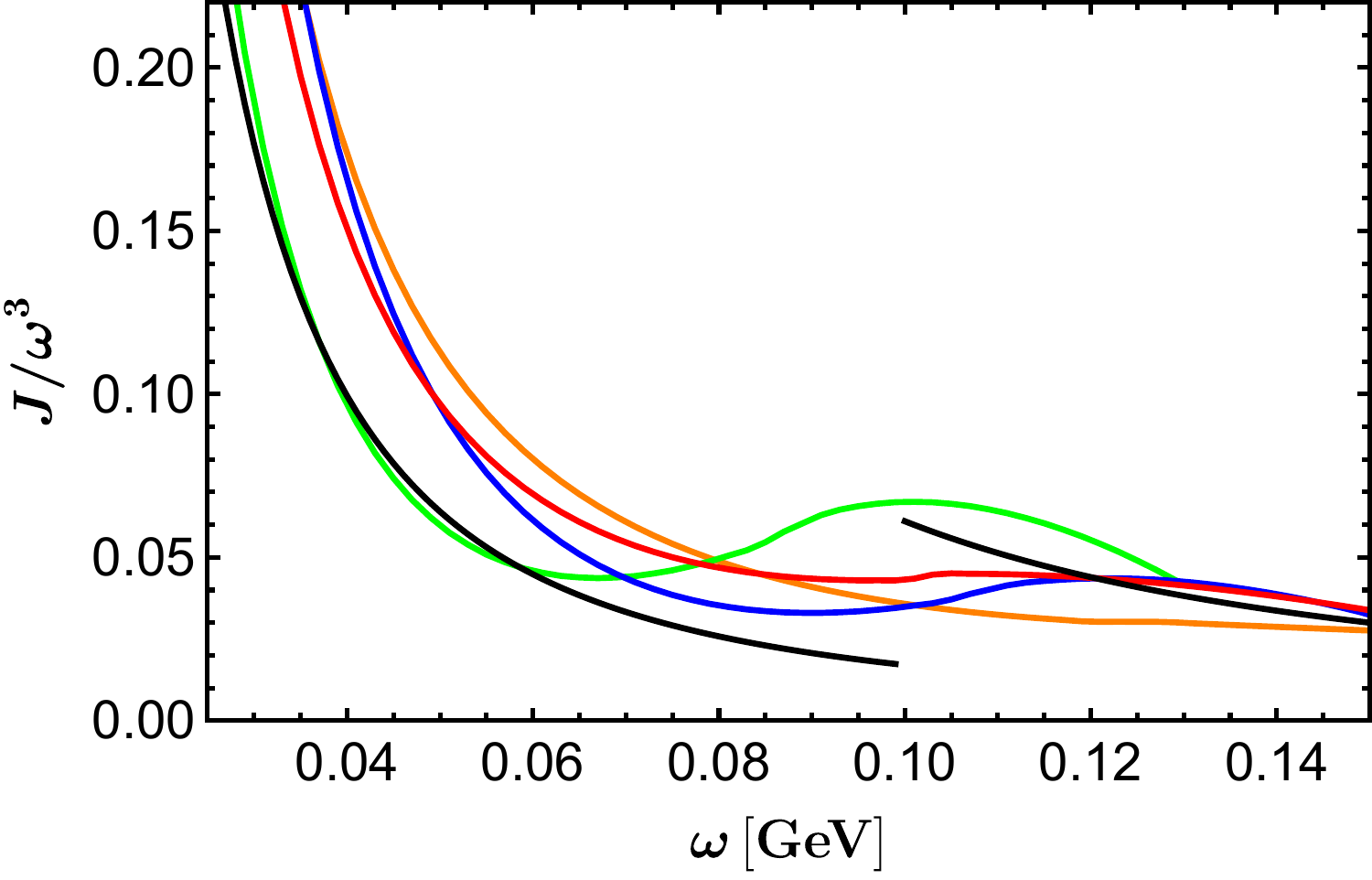}\\
  \caption{ The angular momentum $J$ (left) and the dimensionless combination $J/\omega^3$ (right) as a function of $\omega$. The $\omega$ dependence of $J/\omega^3$ develops a peak structure as approaching a CEP from the crossover region. We have taken $\ell=1\,\text{GeV}^{-1}$.}\label{fig7}
\end{figure}

In Fig.~\ref{fig7}, we show the behavior of angular momentum density $J$ as a function of angular velocity $\omega$. It has been estimated that the size of QGP is about $4-8\,\text{fm}$ at RHIC and is about $6-11\,\text{fm}$ at LHC \cite{Zhang:2001vk}. Therefore, we set $\ell=1\, \text{GeV}^{-1}$ as an estimation and the corresponding angular velocity is $0-1\, \text{GeV}$. We choose representative points on the $\hat{T}-\hat{\mu}_B$ plane.  The orange, red, blue, and green lines represent the points in the crossover region, while the black line corresponds to the one for the first-order phase transition. One can find that $J$ is linear in small $\omega $ in both the hadronic and the QGP phases, which is reminiscent of  the classical formula $J=\rho_m \ell^2\omega$ with $\rho_m$ the mass density. This behavior can be understood using~\eqref{eqJ} from which the small $\omega$ expansion yields
\begin{equation}
J=(\epsilon+P)\ell^2\omega+\mathcal{O}(\omega^3)\,.
\end{equation}
In the non-relativistic limit $\epsilon+P\approx\rho_m$, one immediately recovers the above classical relation between $J$ and $\omega$. One also notes that  the black line suddenly jumps at the first-order transition point. 
An interesting feature arises if one considers the behavior of dimensionless angular momentum density $J/\omega^3$ versus $\omega$, see the right panel of Fig.~\ref{fig7}. As approaching the CEP from the crossover region, there is a peak structure developing near the pseudo-transition point, as shown by the green line. In contrast, referring to the blue and red lines of Fig.~\ref{fig7}, one finds that the peak structure gradually vanishes when the pseudo-transition is away from the CEP. It thus suggests that such peak structure could be a representative signal of the CEP when increasing the angular velocity in experiments.

\subsection{Phase diagram under rotation}
\begin{figure}
    \centering
    \includegraphics[width=0.6\textwidth]{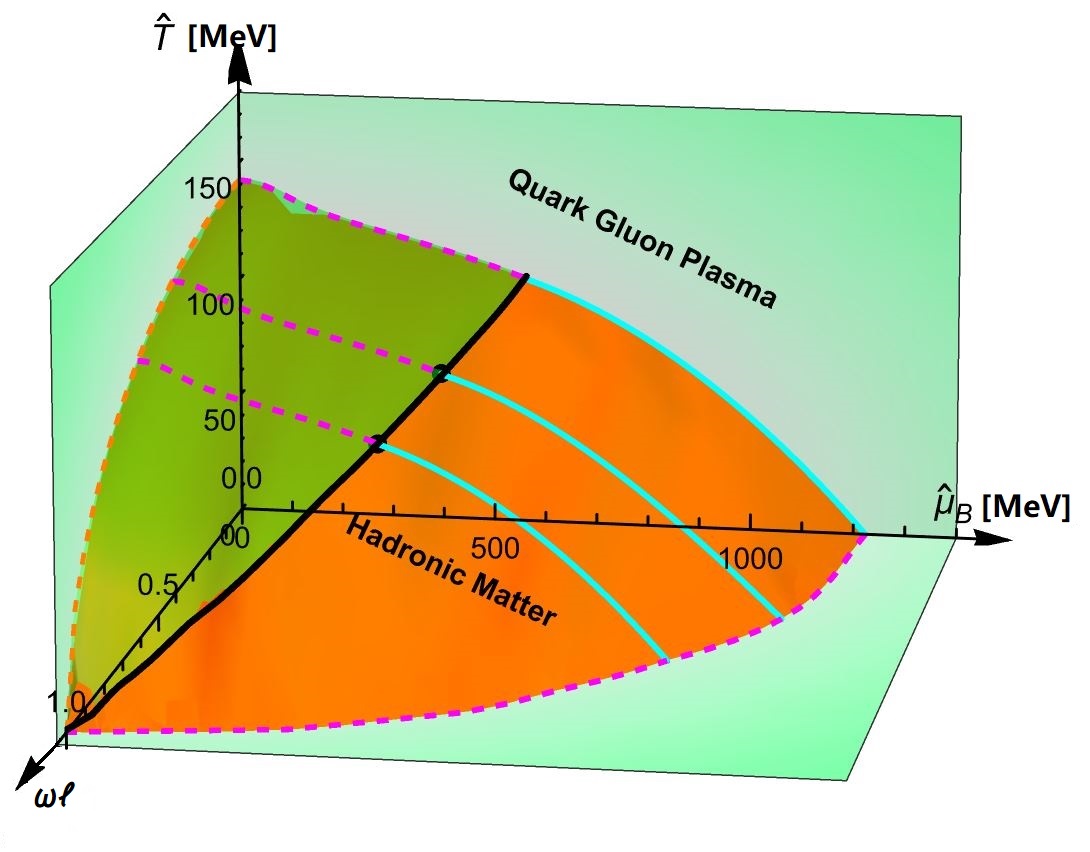}
    \caption{The phase diagram of rotating QCD matter in our 2+1 flavor model. The orange and green surfaces show the phase boundaries for the first-order phase transition and crossover, respectively. The black line denotes the location of the CEP.}
    \label{fig1}
\end{figure}

We show the full phase diagram of 2+1 flavor QCD matter in terms of $\hat{T}$, $\hat{\mu}_B$ and $\omega$ in Fig.~\ref{fig1}. The black solid line denotes the location of CEP. For each angular velocity $\omega$, the smooth crossover between the hadronic phase of color-neutral bound states at low $\hat{T}$ and small $\hat{\mu}_B$, and the QGP at the high $\hat{T}$ and large $\hat{\mu}_B$ changes into the first-order transition for higher chemical potential. It is clear that both the phase transition temperature and chemical potential decrease with increasing angular velocity, which agrees qualitatively with the results of \cite{Wang:2018sur, Chen:2020ath, Chen:2021aiq, Braga:2022yfe, Yadav:2022qcl}.

More precisely, in the $\hat{T}-\hat{\mu}_B$ plane, the position of CEP will slowly shift to the lower left of $\hat{T}-\hat{\mu}_B$ plane as the angular velocity increases. In other words, increasing angular velocity reduces the critical temperature $\hat{T}_{CEP}$ and chemical potential $(\hat{\mu}_B)_{CEP}$. Moreover, for a fixed baryon chemical potential $\hat{\mu}_B$, there is also a nontrivial CEP on the $\hat{T}-\omega$ plane when $0<\hat{\mu}_B<\mu_C$ where $\mu_C=558\,\text{MeV}$ is the baryon chemical potential of CEP at $\omega=0$ \cite{Cai:2022omk}. For  $\hat{\mu}_B> \mu_C$, the CEP on the $\hat{T}-\omega$ plane will vanish, for which only the first-order phase transition exists.

\begin{figure}
  \centering
  \includegraphics[width=0.7\textwidth]{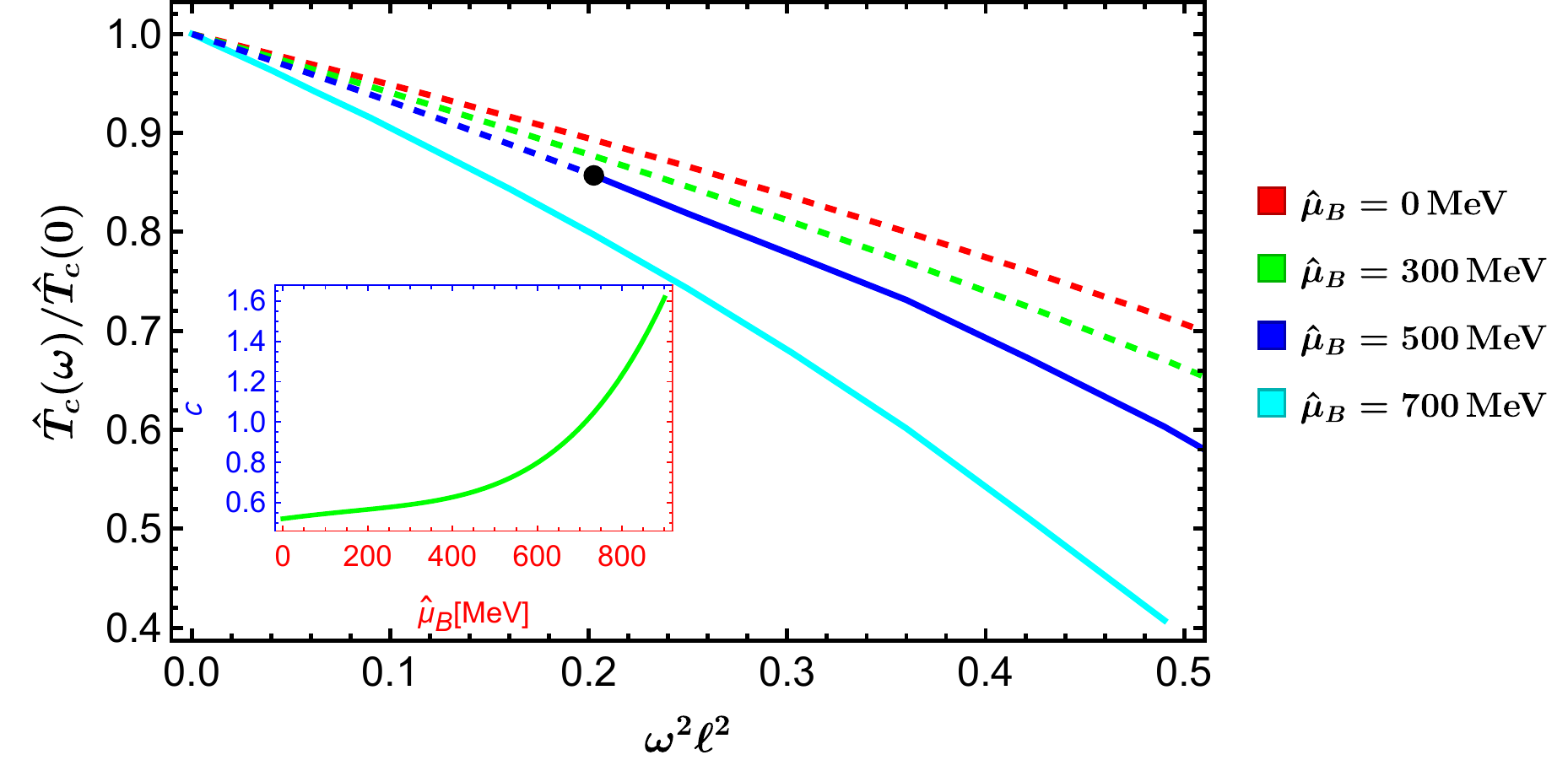}\\
  \caption{The critical temperature $\hat{T}_c$ as a function of $\omega$ for different $\hat{\mu}_B$. For small $\omega$, $\hat{T}_c(\omega)/\hat{T}_c(0)\approx1-c\,\omega^2$ where the constant $c$ increases as $\hat{\mu}_B$ is increased (see the Insert).}\label{fig2}
\end{figure}

The lattice QCD simulation \cite{Braguta:2021ucr} has suggested that the influence of rotation on fermions and gluons should be opposite. In gluon-dynamics, the confinement-deconfinement phase transition temperature increases with the angular velocity. Moreover, the phase transition temperature $\hat{T}_c$ is determined from the peak of the Polyakov loop susceptibility and the ratio $\hat{T}_c(\omega)/\hat{T}_c(0)=1+C_2\,\omega^2$ with $C_2$ is a positive constant. For the $N_f=2$ Wilson fermions, it was argued that rotating fermions attempt to decrease the phase transition temperature. Unfortunately, the detailed behavior for the critical temperature as a function of angular velocity was not presented \cite{Braguta:2021ucr}. In the present work, we find that the phase transition temperature decreases with the increase of angular velocity. More precisely, when $\omega$ is small, it follows that $\hat{T}_c(\omega)/\hat{T}_c(0)\approx1-c\,\omega^2$ with $c$ a positive constant. The value of $c$ depends on $\hat{\mu}_B$, since $\hat{\mu}_B$ is also related to $\omega$. In particular, at $\hat{\mu}_B=0$, one has
\begin{equation}
    \hat{T}_c(\omega)=\frac{\hat{T}_c(0)}{\sqrt{1-\omega^2\ell^2}} \approx (1-\frac{\omega^2\ell^2}{2})\hat{T}_c(0)\,.
\end{equation}
The critical temperature as a function of $\omega^2$ for various values of $\hat{\mu}_B$ is shown in Fig.~\ref{fig2}. The dashed lines show the phase transition line in the crossover region, determined by the minimum of {$\hat{c}_s^2$}, while the solid lines are the first-order transition lines. The value of $c$ increases with the increase of $\hat{\mu}_B$. 

\begin{figure}
  \centering
  \includegraphics[width=0.32\textwidth]{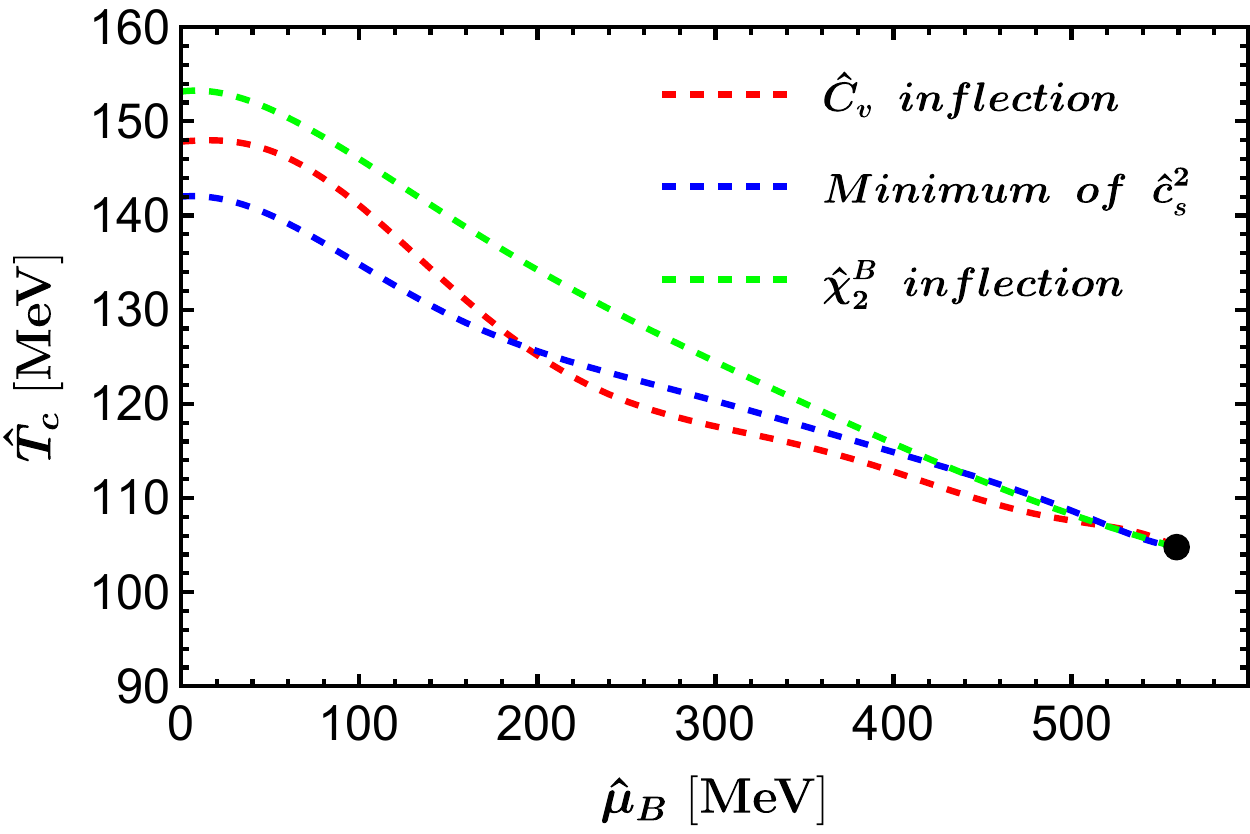}
  \includegraphics[width=0.32\textwidth]{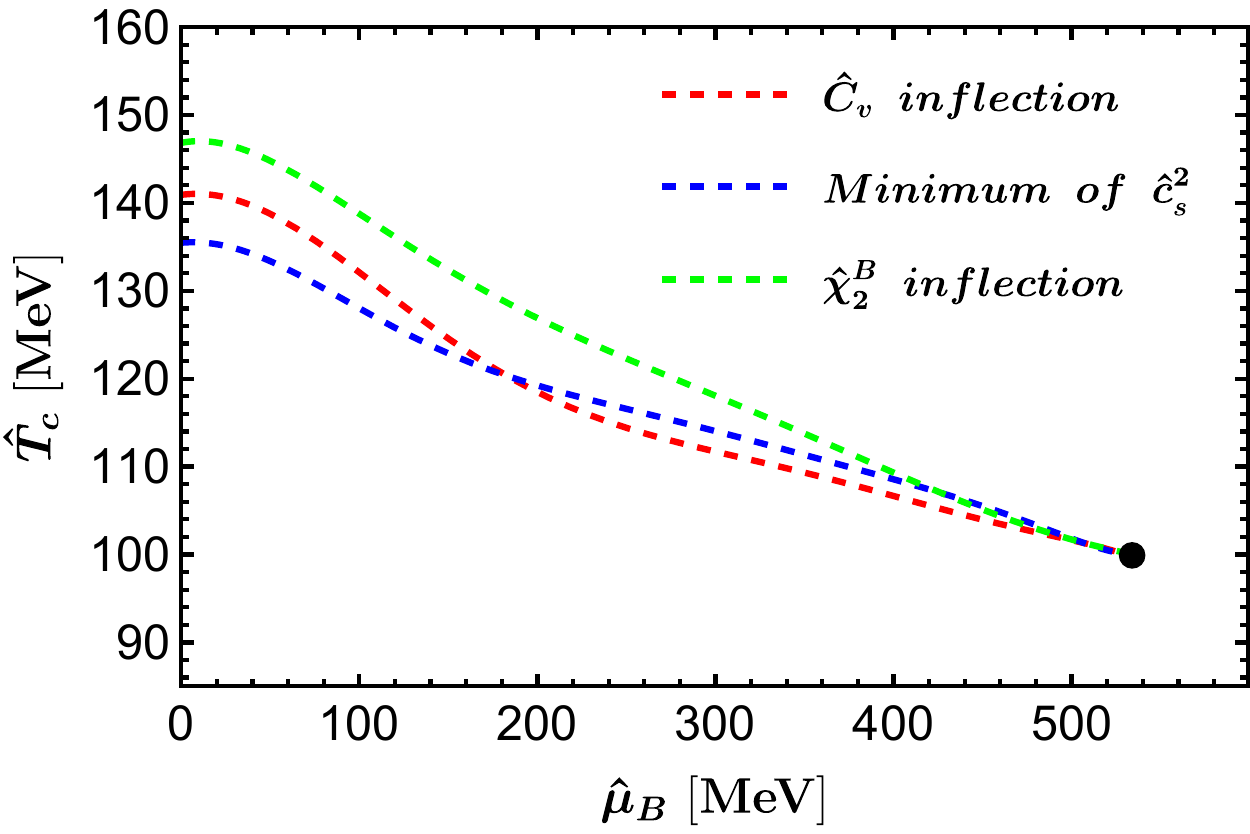}
  \includegraphics[width=0.32\textwidth]{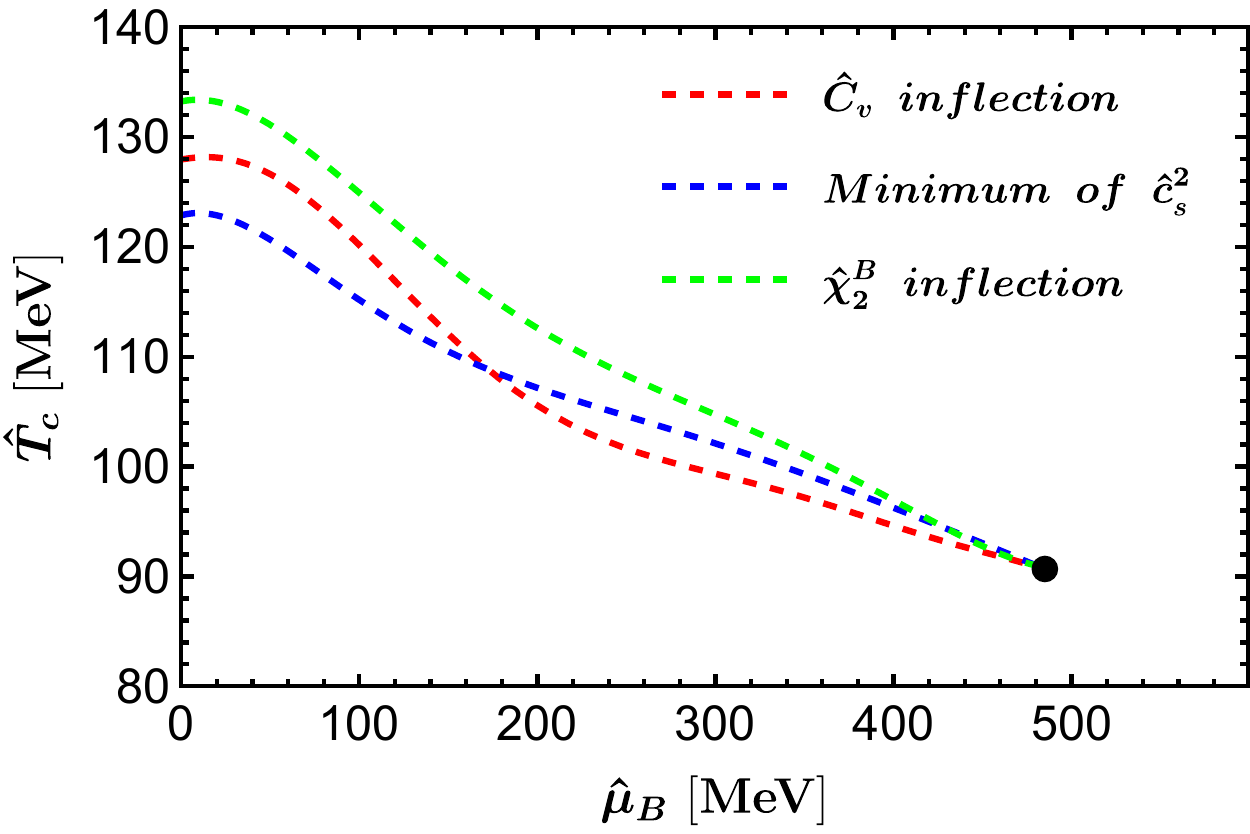}\\
  \caption{The phase boundaries determined by the inflection of $\hat{C}_v$ (red), the minimum of {$\hat{c}_s^2$} (blue) and the inflection of $\hat{\chi}_2^B$ (green) in crossover region. Three plots correspond to $\omega\ell=0,~0.3~\text{and}~0.5$ from left to right.}\label{fig3}
  \end{figure}

There is no unique way to determine the transition temperature in the literature for smooth crossover. We have applied the minimum of {$\hat{c}_s^2$} to define the pseudo-transition temperature. Other quantities can be also used to characterize the drastic change in degrees of freedom between the QGP and the hadron phase. In Fig.~\ref{fig3}, we show the locations of crossover determined by the inflection of $\hat{C}_v$, the minimum of {$\hat{c}_s^2$} and the inflection of $\hat{\chi}_2^B$. Although the different criteria give different results, they agree with each other qualitatively.

\subsection{Pure gluon}

Before ending this section, we would like to show the results for the pure gluon system at vanishing chemical potential. It has been shown that there is a strong first-order confinement/deconfinement phase transition at $T_c=276.5\,\text{MeV}$ \cite{He:2022toapp} without rotation.

\begin{figure}
  \centering
  \includegraphics[width=0.55\textwidth]{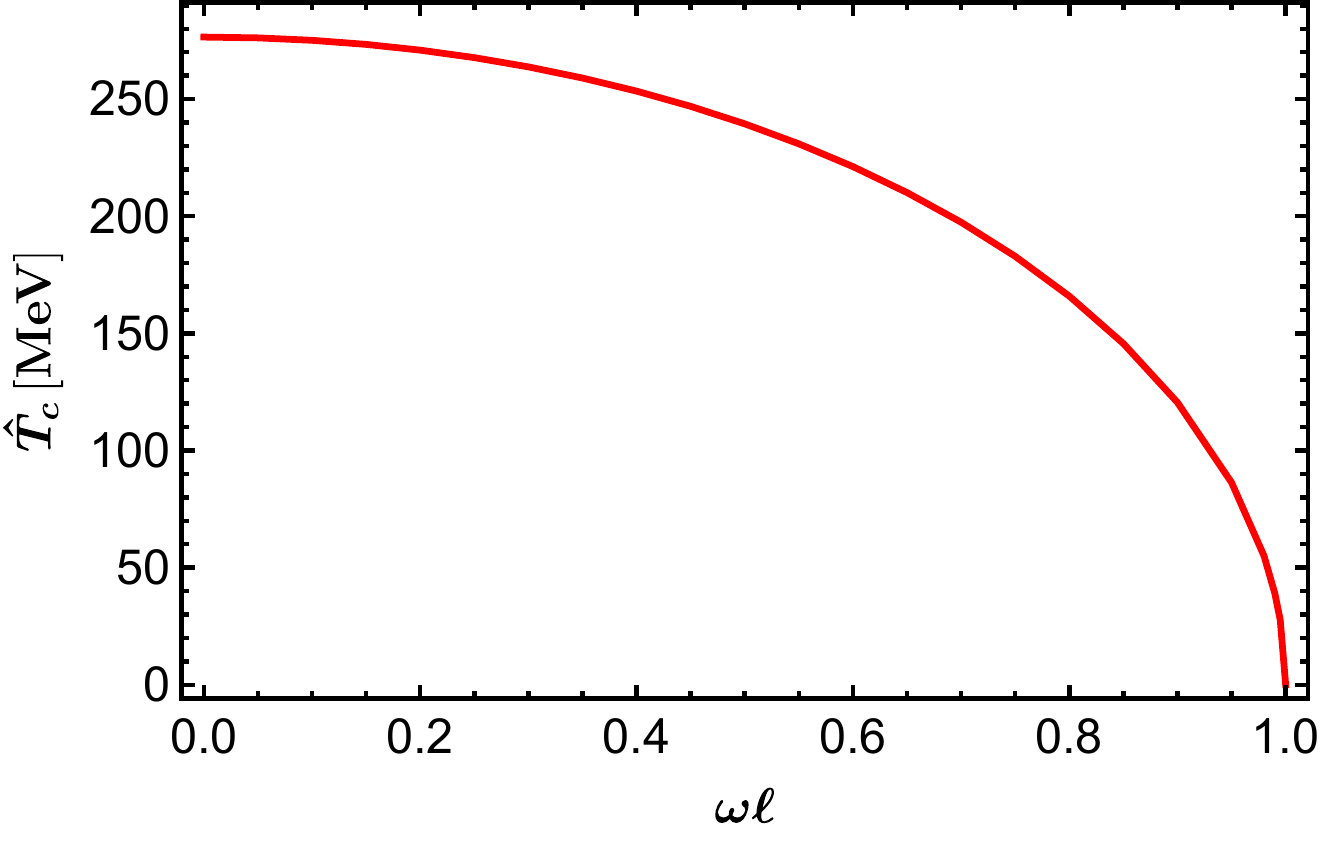}\\
  \caption{The $\hat{T}-\omega$ phase diagram of the pure gluon system. The critical temperature of the first-order phase transition decreases as $\omega$ is increased.}\label{fig9}
\end{figure}

The phase diagram of the temperature and angular velocity is shown in Fig.~\ref{fig9}. Since the chemical potential is vanishing, the critical temperature as a function of $\omega$ can be obtained analytically, i.e., $\hat{T}_c(\omega)=T_c \sqrt{1-\omega^2\ell^2}$ (the red curve in Fig.~\ref{fig9}). Note that the angular velocity is normalized so that the transition temperature will decrease to zero as $\omega\ell \to 1$.

\begin{figure}
  \centering
  \includegraphics[width=0.49\textwidth]{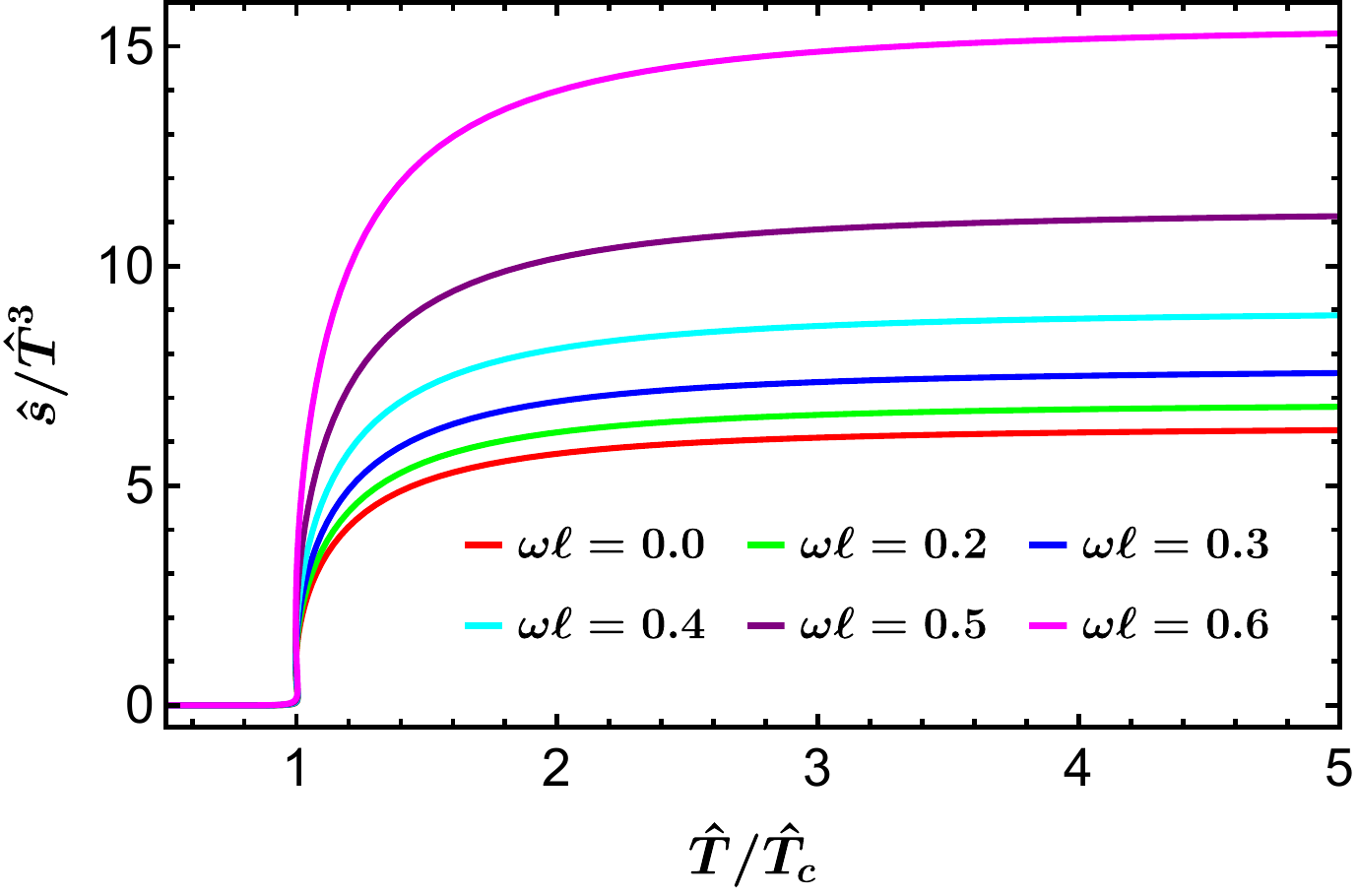}
  \includegraphics[width=0.49\textwidth]{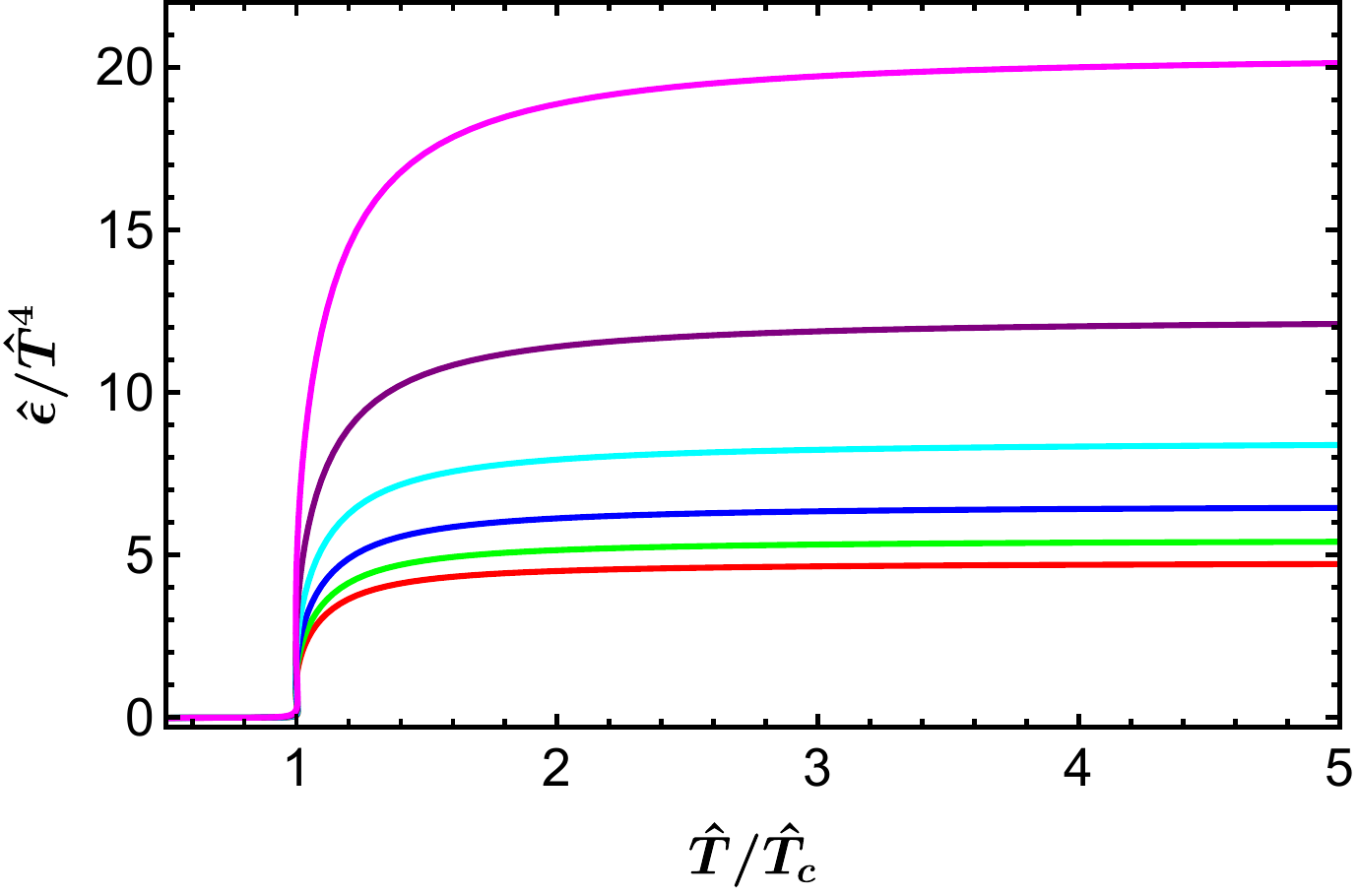}\\
  \includegraphics[width=0.49\textwidth]{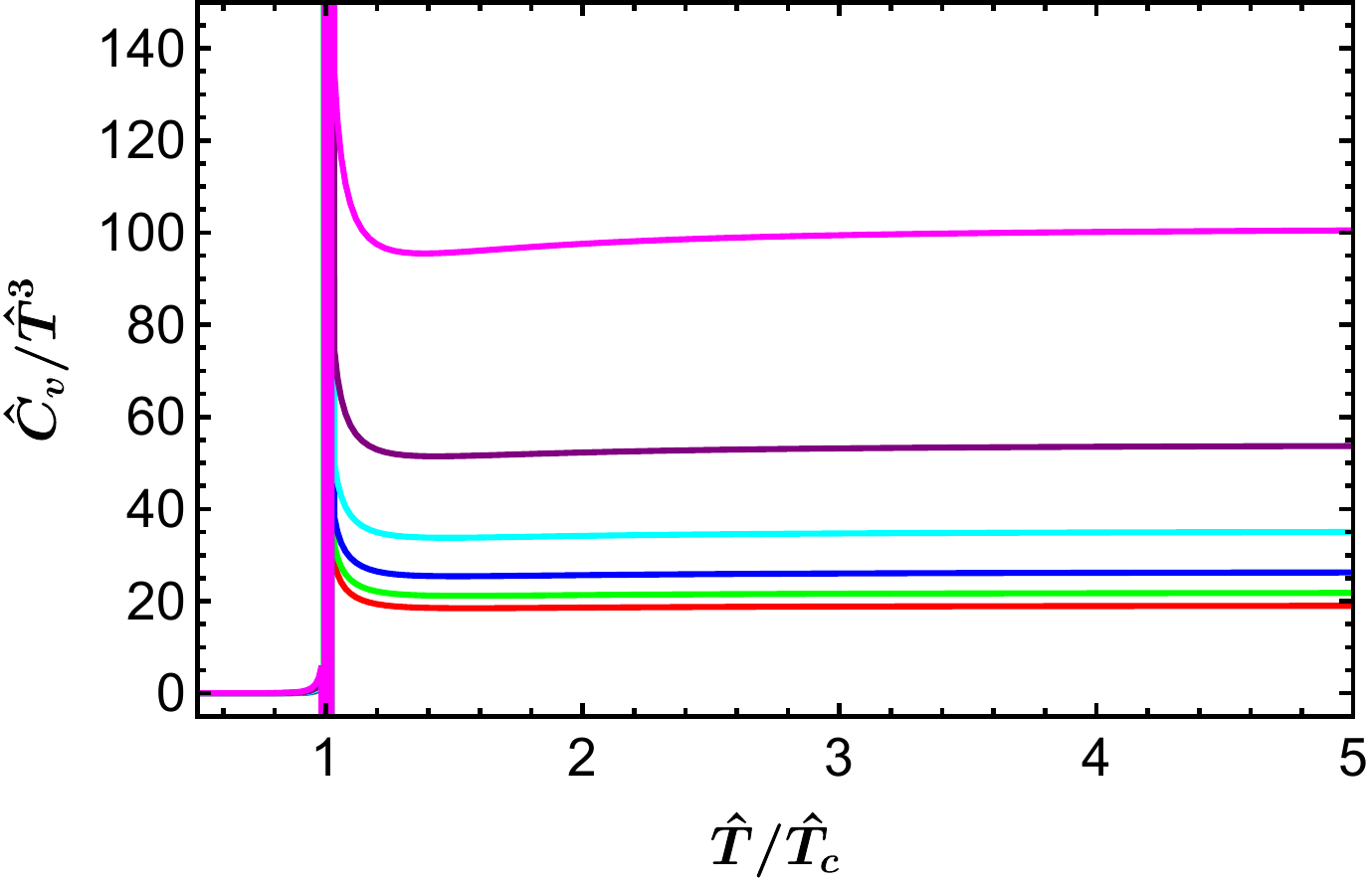}
  \includegraphics[width=0.49\textwidth]{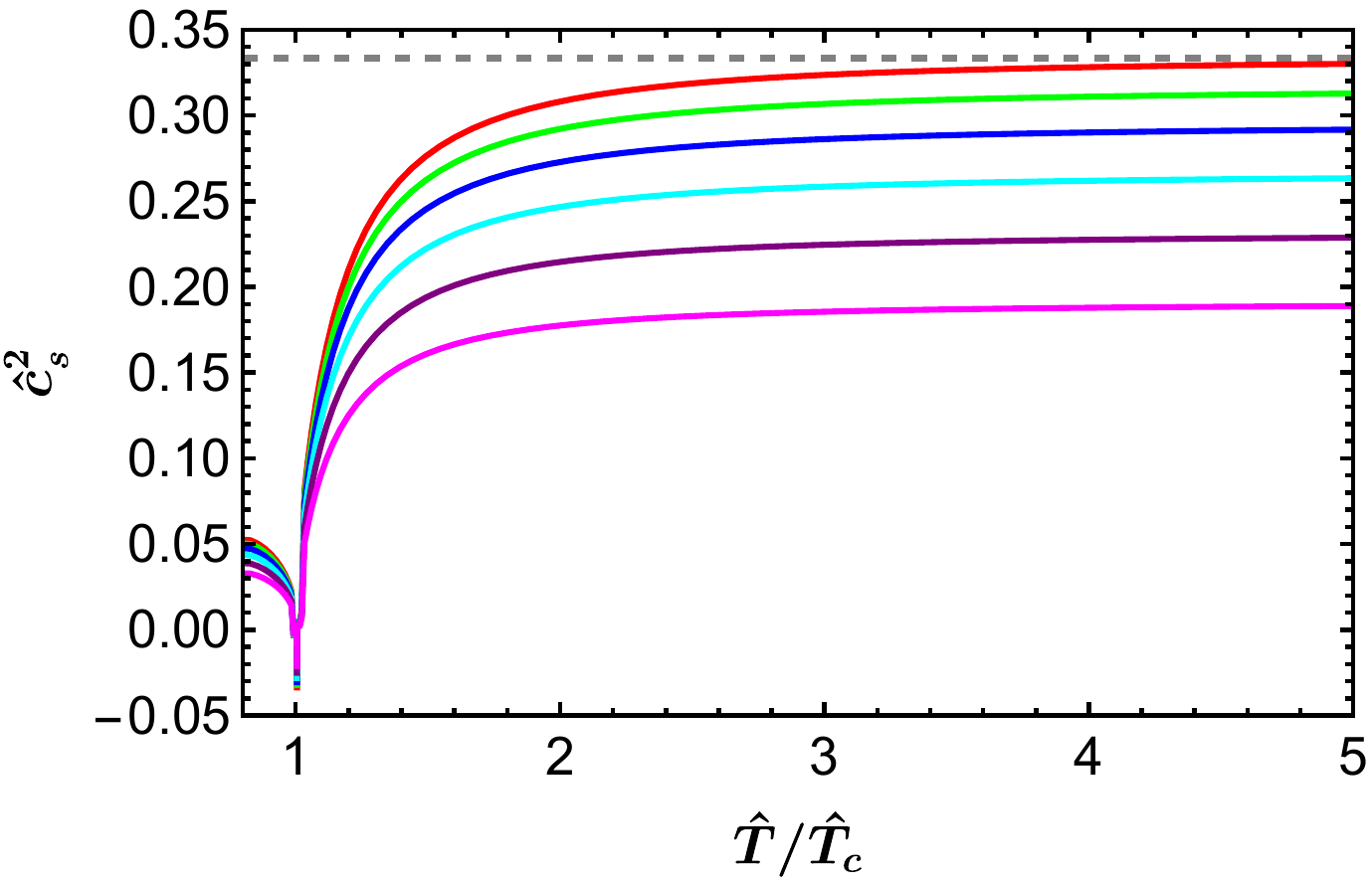}\\
  \caption{Various thermodynamic quantities in function of normalized temperature $\hat{T}/\hat{T}_c$ of the pure gluon system at different angular velocities. With the increase of $\omega$, $\hat{s}/\hat{T}^3$, $\hat{\epsilon}/\hat{T}^4$ and $\hat{C}_v/\hat{T}^3$ are enhanced, while  $\hat{c}_s^2$ is suppressed.}\label{fig10}
\end{figure}

\begin{figure}
  \centering
  \includegraphics[width=0.49\textwidth]{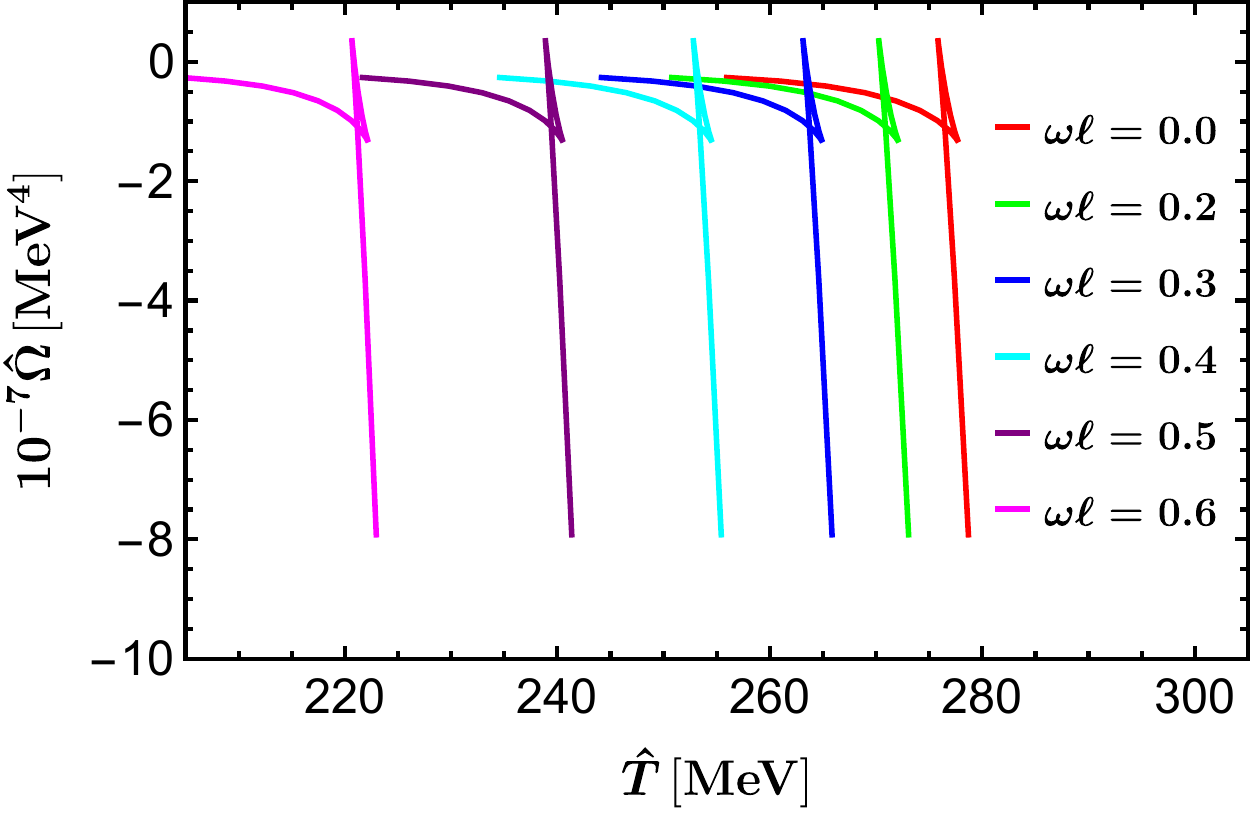}
  \includegraphics[width=0.49\textwidth]{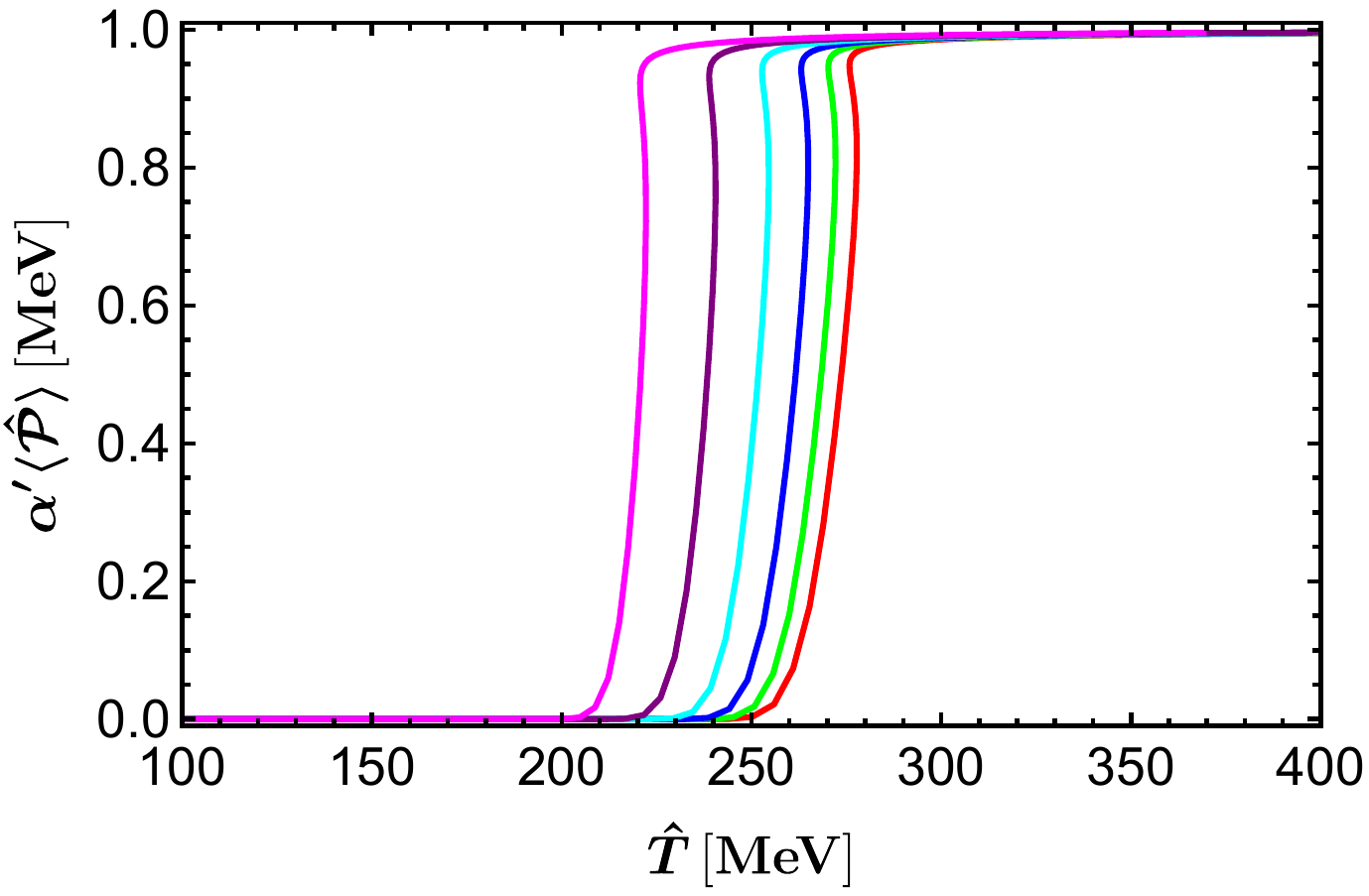}\\
  \caption{The free energy $\hat{\Omega}$ (left) and the Polyakov loop $\langle \hat{\mathcal{P}} \rangle$ at different $\omega$, suggesting a first-order deconfinement phase transition. }\label{fig12}
\end{figure}

Various thermodynamic quantities as a function of normalized temperature $\hat{T}/\hat{T}_c$ are presented in Fig.~\ref{fig10}, including the entropy density $\hat{s}$, energy density $\hat{\epsilon}$, specific heat $\hat{C}_v$ and squared speed of sound $\hat{c}_s^2$ . While $\hat{s}/\hat{T}^3$, $\hat{\epsilon}/\hat{T}^4$ and $\hat{C}_v/\hat{T}^3$ are enhanced by increasing $\omega$, $\hat{c}_s^2$ is suppressed by the rotation. Fig.~\ref{fig12} shows the behaviors of the free energy $\hat{\Omega}$ and the Polyakov loop $ \langle \hat{\mathcal{P}} \rangle$ with respect to $\omega$. The sharp of $\hat{\Omega}(\hat{T})$ does not change when varying angular velocity $\omega$. As $\omega$ increases, $\hat{\Omega}$ shifts to a lower temperature. As shown in the right panel of Fig.~\ref{fig12}, the Polyakov loop also shifts to a lower temperature with the increase of $\omega$. Note that the Polyakov loop is a good order parameter for the deconfinement for the pure gluon system, and its expectation value vanishes at the low-temperature phase. A first-order confinement/deconfinement is manifesting.

\begin{figure}
  \centering
  \includegraphics[width=0.49\textwidth]{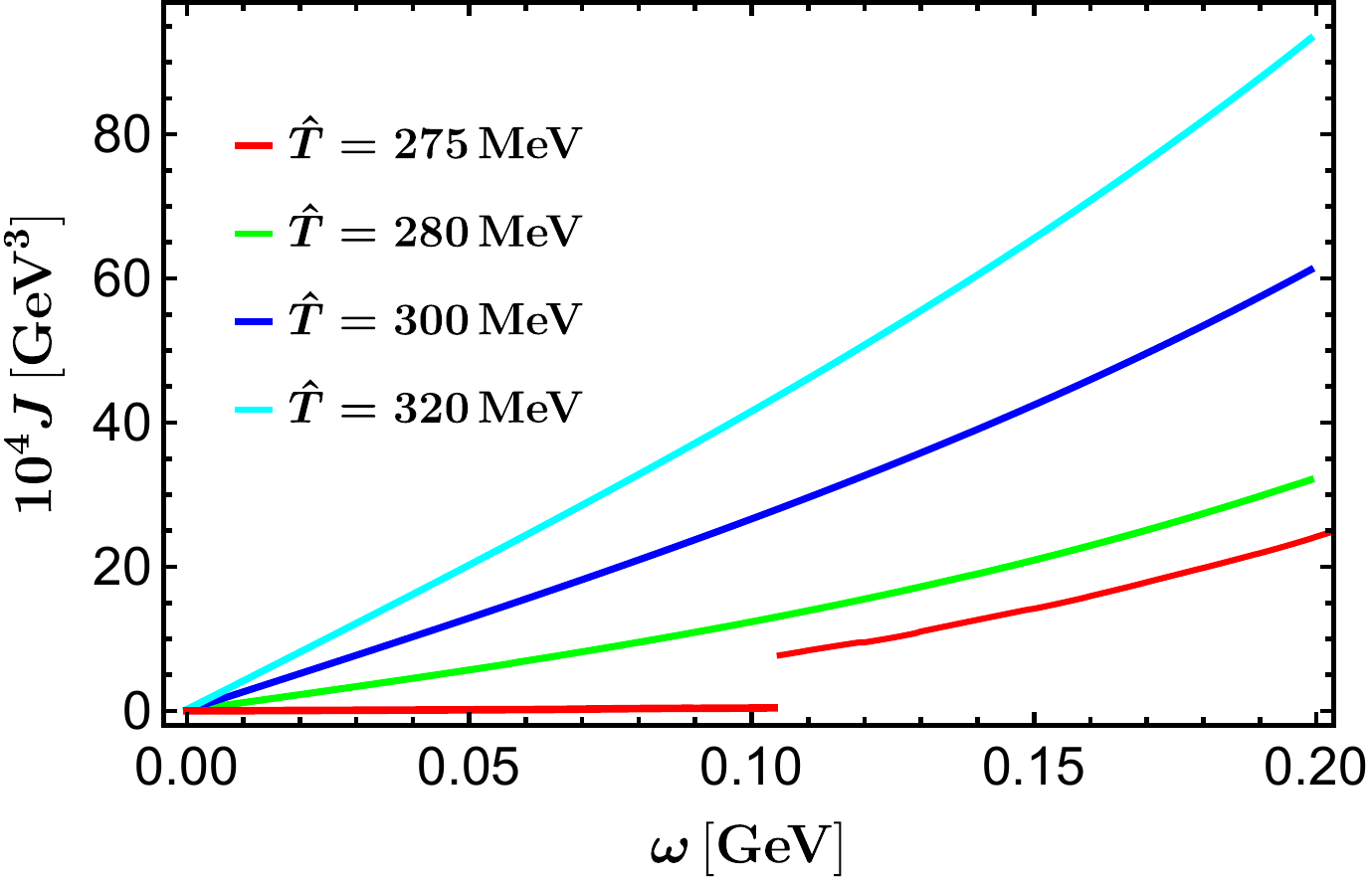}
  \includegraphics[width=0.49\textwidth]{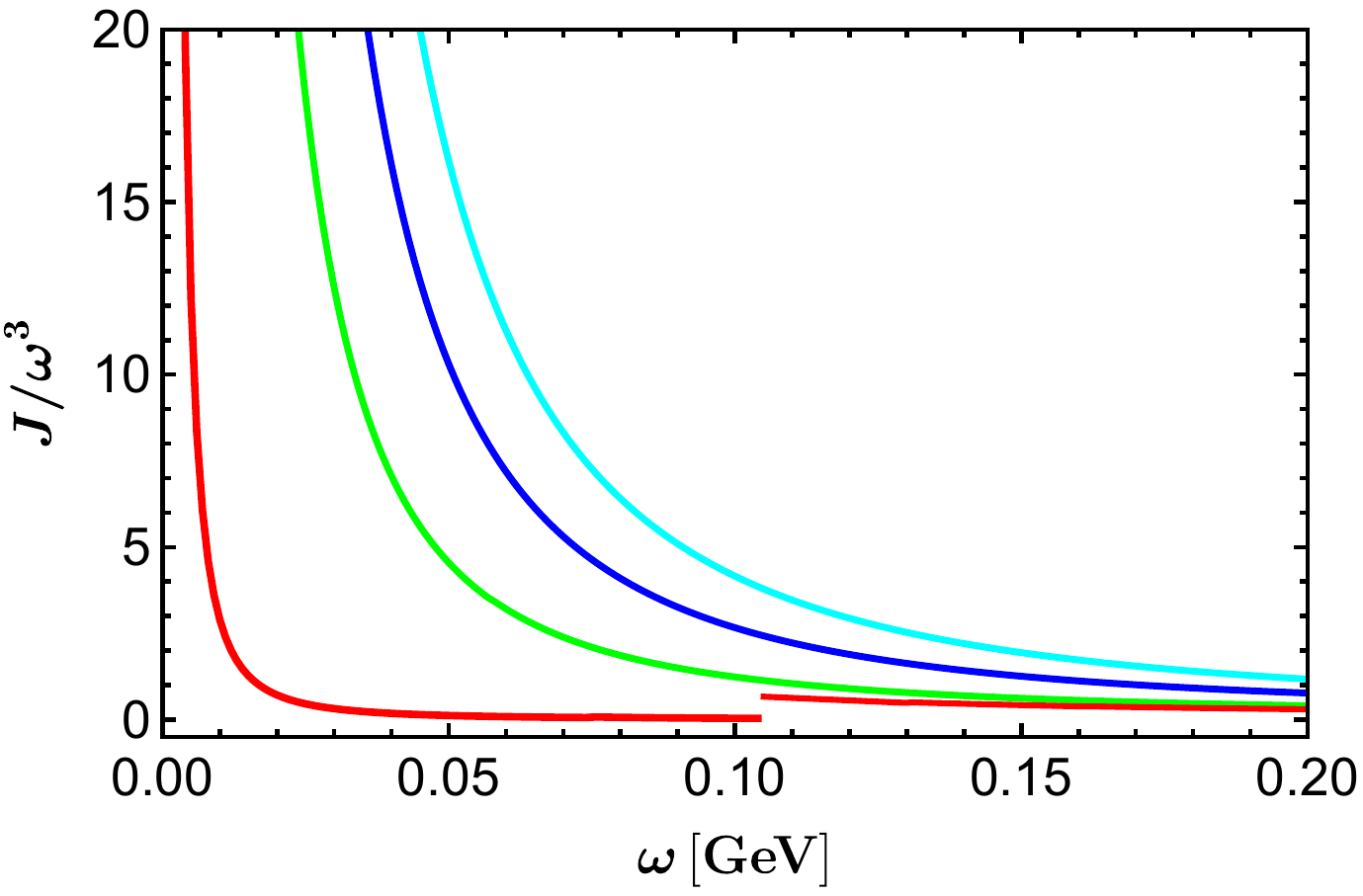}\\
  \caption{The angular momentum $J$ (left) and $J/\omega^3$ (right) in function of $\omega$ for different temperatures. We have taken $\ell=1\,\text{GeV}^{-1}$. }\label{fig13}
\end{figure}

We show the behavior of angular momentum density $J$ as a function of angular velocity $\omega$ in the left panel Fig.~\ref{fig13}. For a fixed temperature below $T_c$, the angular momentum $J$ has a discontinuous jump at a particular value of $\omega$, signaling the first-order transition shown in Fig.~\ref{fig9}. The case with $\hat{T}>T_c$ increases smoothly as $\omega$ is increased. The dimensionless angular momentum density $J/\omega^3$ as a function of $\omega$ is given in the right panel of Fig.~\ref{fig13}. One finds that $J/\omega^3$ decrease monotonically with the increase of $\omega$ in each case except the discontinuity at the phase boundary. Since there is no CEP, one does not see the peak structure shown in the right panel of Fig.~\ref{fig7}.

\section{Summary and discussion}\label{sec:04summary}
We have considered the rotation effect on the phase structure of $2+1$ flavor and pure gluon systems. Without rotation, it has been shown that there is a first-order phase transition at $T_ c=276.5\,\text{MeV}, \mu_B=0\,\text{MeV}$ for pure gluon system, while for 2+1 flavor system, there is a CEP located at ($T_c=105\,\text{MeV}$, $\mu_B=555\,\text{MeV}$).
The local Lorentz boost is introduced in the bulk background as an approximate description of QCD matter under rotation.

We have obtained various thermodynamic quantities using holographic renormalization in a self-consistent way. In particular, we can directly get an explicit expression for the angular momentum $J$~\eqref{eqJ} and check the thermodynamic relation~\eqref{eqrelation}. We have found that the $\hat{T}$ scaled entropy density, energy density, specific heat, and baryon susceptibility will be enhanced by angular velocity (see Figs.~\ref{fig4} and~\ref{fig10}). At the same time, the speed of sound is suppressed under rotation, as shown in Fig.~\ref{fig5p}. Furthermore, the angular momentum monotonically increases for both systems as the angular velocity $\omega$ increases. At small $\omega$, one has $J=(\epsilon+P)\ell^2\omega$ which, in the non-relativistic lime, reduces to the classical formula $J=\rho_m \ell^2\omega$ with $\rho_m$ the mass density. Interestingly, as approaching the CEP from the crossover region, the dimensionless angular momentum $J/\omega^3$ versus $\omega$ develops a peak (see the right panel of Fig.~\ref{fig7}). It suggests that such peak structure could be an effective signal of the CEP when increasing the angular velocity in experiments. Understanding this peak might be helpful to enable the experimental search for the CEP.

For the $N_f=2+1$ system, the transition is always a crossover when $\hat{\mu}_B=0$. For finite $\hat{\mu}_B$, as shown in Fig.~\ref{fig6}, the transition would turn to be first-order at large $\omega$, yielding a CEP between the first-order line and crossover region. The location of the CEP has been shown to shift towards low temperature and small chemical potential as $\omega$ is increased. We have constructed the complete phase diagram of 2+1 flavor QCD matter in terms of $\hat{T}$, $\hat{\mu}_B$ and $\omega$, see Fig.~\ref{fig1}. The CEP will shrink quickly to the point $(0,0,1)$  on  the $\omega\ell$ axis as $\omega\ell\rightarrow1$. With a fixed $\omega$, the temperature of CEP will also decrease with $\hat{\mu}_B$. For the pure gluon system with $\hat{\mu}_B=0$, there is a first-order confinement/deconfinement phase transition as shown in Fig.~\ref{fig12}. The critical temperature as a function of $\omega$ is given as $\hat{T}_c(\omega)\sim \hat{T}_c(0)\sqrt{1-\omega^2\ell^2}$ (see Fig.~\ref{fig9}).

Dedicating to a precise characterization of QCD matter under rotation, particularly the properties and differences of the QGP and hadronic phases along the first-order phase transition in 2+1 flavor QCD matter, is an interesting direction for further study. Although some interesting results have been obtained, the present study is just a preliminary approximation of the rotation effect. It will be desirable to consider more realistic rotating QCD matter for which the distribution of the rotating system would depend on the distance to the rotating axis. {Recently, the numerical simulation of Euclidean SU(3) Yang-Mills plasma rotating with a high imaginary angular frequency suggested the emergence of a spatially inhomogeneous confining-deconfining phase transition~\cite{Chernodub:2022veq}. Holographic realization of an inhomogeneous structure of QCD matter at rotation would be a very interesting step forward.} In this case, one must solve partial differential equations to get a more realistic configuration. Moreover, the present research should be embedded in a more general and multidimensional view of the QCD phase diagram, including magnetic field, isospin, etc. We shall leave them to future work.

\section*{Acknowledgements}
We would like to thank Maxim N. Chernodub, Yu Guo, Hai-cang Ren, and Xin-li Sheng for their useful discussions. This work is partly supported by the National Natural Science Foundation of China (NSFC) under Grant Nos. 11735007, 11890711, 12075298, 11890710, 12275104, 11947233, 12075101, 12235016, and 12122513. S. H. also would like to appreciate the financial support from Jilin University and the Max Planck Partner group. L.L. also appreciates the Chinese Academy of Sciences Project for Young Scientists in Basic Research YSBR-006.
	
	
\providecommand{\href}[2]{#2}\begingroup\raggedright\endgroup

\end{document}